\documentclass[tighten,twocolumn]{aastex62}

\usepackage{color}

\usepackage{graphicx}
\usepackage{rotating}
\usepackage{float}

\usepackage{enumitem}
\usepackage{amssymb}

\begin{document}

\shorttitle{Properties of the Umbral Filament in AR NOAA 12529}
\shortauthors{Guglielmino et al.}

\title{Properties of the Umbral Filament Observed in Active Region NOAA 12529}

\author{Salvo L. Guglielmino}
\affiliation{Dipartimento di Fisica e Astronomia ``Ettore Majorana'' -- Sezione Astrofisica, Universit\`{a} degli Studi di Catania, Via S.~Sofia 78, I-95123 Catania, Italy}

\author{Paolo Romano}
\affiliation{INAF -- Osservatorio Astrofisico di Catania,
	Via S.~Sofia 78, I-95123 Catania, Italy}

\author{Basilio Ruiz~Cobo}
\affiliation{IAC -- Instituto de Astrof\'isica de Canarias, C/ V\'ia L\'actea s/n, E-38200, La Laguna, Tenerife, Spain}
\affiliation{ULL -- Departamento de Astrof\'isica, Univ.~de La Laguna, E-38205, La Laguna, Tenerife, Spain}

\author{Francesca Zuccarello}
\affiliation{Dipartimento di Fisica e Astronomia ``Ettore Majorana'' -- Sezione Astrofisica, Universit\`{a} degli Studi di Catania, Via S.~Sofia 78, I-95123 Catania, Italy}

\author{Mariarita Murabito}
\altaffiliation{currently at INAF -- Osservatorio Astronomico di Roma, \\
	via Frascati 33, I-00078 Monte Porzio Catone, Italy}
\affiliation{Dipartimento di Fisica e Astronomia ``Ettore Majorana'' -- Sezione Astrofisica, Universit\`{a} degli Studi di Catania, Via S.~Sofia 78, I-95123 Catania, Italy}

\correspondingauthor{Salvo L. Guglielmino}
\email{salvatore.guglielmino@inaf.it}

\begin{abstract}

Recent observations of the solar photosphere revealed the presence of elongated filamentary bright structures inside sunspot umbrae, called umbral filaments (UFs). These features differ in morphology, magnetic configuration, and evolution from light bridges that are usually observed to intrude in sunspots. To characterize an UF observed in the umbra of the giant leading sunspot of active region NOAA 12529, we analyze high-resolution observations taken in the photosphere with the spectropolarimeter aboard the \textit{Hinode} satellite and in the upper chromosphere and transition region with the \textit{IRIS} telescope. The results of this analysis definitely rule out the hypothesis that the UF might be a kind of light bridge. In fact, we find no field-free or low-field strength region cospatial to the UF. Conversely, we recognize the presence of a strong horizontal field larger than 2500~G, a significant portion of the UF with opposite polarity with respect to the surroundings, and filaments in the upper atmospheric layers corresponding to the UF in the photosphere. These findings suggest that this structure is the photospheric manifestation of a flux rope hanging above the sunspot and forming penumbral-like filaments within the umbra via magneto-convection. This reinforces a previously proposed scenario.

\end{abstract}

\keywords{sunspots --- Sun: photosphere --- Sun: chromosphere --- Sun: transition region --- Sun: UV radiation --- Sun: magnetic fields}

\section{Introduction}

The advent of high resolution observations during these last decades has led to an overwhelming progress in the understanding of the interplay between plasma and magnetic fields that populate the solar atmosphere. In this respect, data from the 1-m class aperture ground-based facilities and from space-borne free-seeing telescopes have provided new insights into the fine structure of sunspots, which are the most prominent manifestation of solar activity \citep{Solanki:03, Schlichenmaier:09}. An aspect that is currently under special investigations is the magnetic coupling between the photosphere and the upper atmospheric layers above sunspots and their sub-structures \citep{BI:11}.

Bright features that intrude sunspots from the leading edge of penumbra into the umbra are known as light bridges (LBs). They are often observed during the decay phase of the sunspot evolution \citep{Vazquez:73}, albeit some of them are noticed during the assembly phase of sunspots as well \citep[see also][and references therein]{Falco:16,Felipe:16}. \citet{Sobotka:97} proposed two morphological classifications of LBs into strong or faint, whether they separate the umbra into distinct umbral cores, and into granular or filamentary LBs, depending on the pattern observed in their internal structure. \citet{Thomas:04}, for their part, distinguished between segmented and unsegmented LBs. Unsegmented LBs resemble elongated penumbral filaments, without evidence of granules in their interior. Conversely, segmented LBs exhibit bright granular cells that are thought to have the same convective origin as quiet Sun granulation, but differing in size, lifetime, and brightness \citep[see, e.g.,][]{Lagg:14,Falco:17}. Recent observations performed at the GREGOR telescope reported the presence of another different class of LBs, i.e., thick LBs, with small transverse \textsc{Y}-shaped dark lanes similar to dark-cored penumbral filaments connected to the central dark lane \citep{Schlichenmaier:16}.

In this context, \citet{Kleint:13} studied unusual filamentary structures observed within the umbra of the very flare-productive active region (AR) NOAA~11302 preceding sunspot. They are formed by curled filaments that intrude sunspots from the penumbra well into the umbra. Along these structures, counter-Evershed flows are observed at the photospheric level and energy dissipation phenomena occur in the higher atmospheric layers. \citet{Kleint:13} proposed to call these structures, which do not resemble typical LBs in morphology or in evolution, as umbral filaments (UFs). Also \citet{Siu-Tapia:17} discussed on the possible presence of UFs to explain counter-Evershed flows observed in large area of a sunspot penumbra.

In a previous work \citep*[][hereafter Paper~I]{Guglielmino:17}, we have studied an elongated bright structure, being similar to a filamentary LB (FLB), which was observed inside the umbra of the giant preceding sunspot of AR NOAA~12529 (hereafter, AR~12529). Using observations acquired by the Helioseismic and Magnetic Imager \citep[HMI;][]{Scherrer:12} on board \textit{Solar Dynamic Observatory} \citep[SDO;][]{Pesnell:12} satellite, we have found that this intruding feature was characterized by a strong horizontal magnetic field, with a portion of the structure having opposite polarity to that of the hosting sunspot. Moreover, mixed directions of the plasma motions were apparently observed along the main axis of the structure. Comparing these characteristics with those of LBs and taking account for the cospatial structures seen in the chromospheric and coronal layers, this feature was interpreted as an UF, actually being the counterpart of a flux rope that was located in higher layers of the solar atmosphere above the umbra.

In this Paper, we describe high-resolution observations performed by the spectropolarimeter aboard the \textit{Hinode} satellite in the photosphere and by the \textit{IRIS} spacecraft in the upper chromosphere and transition region, at a time close to the central meridian passage of AR~12529. The spectropolarimetric measurements show a strong horizontal field in the UF, which also exhibits a portion with magnetic field of polarity opposite to that of the surrounding umbra, even larger than that observed with the \textit{SDO}/HMI. Moreover, the configuration of the structure seen in \textit{IRIS} observations appears to be highly reminiscent of a flux rope. 

The paper is organized as follows. We describe the observations in Section~2 and report the results in Section~3. We summarize our findings and draw conclusions in Section~4.

\begin{figure}[t]
	\centering
	\includegraphics[trim=70 95 80 85, clip, scale=0.325]{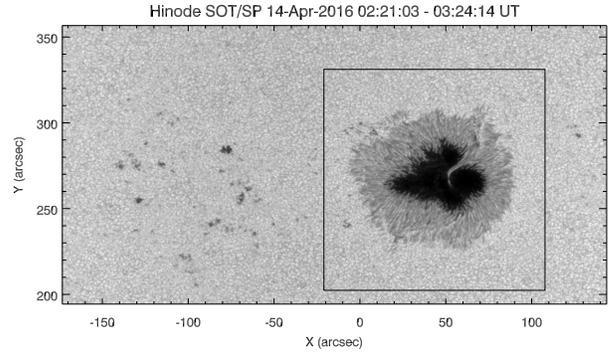}
	\caption{Reconstructed continuum map of AR 12529 built from the \textit{Hinode} SOT/SP scan here analyzed. The box frames the sub-FoV studied in the paper (see the following Figures). Here and in the following maps, the solar north is on the top, and west is at the right. \label{fig:context}}
\end{figure}

\section{Data and analysis}

To analyze the photospheric evolution of the feature during the passage of AR~12529 across the solar disc, we used the continuum filtergrams acquired along the \ion{Fe}{1} line at 617.3~nm by the \textit{SDO}/HMI instrument. These images are part of Space-weather Active Region Patches (SHARPs) data \citep{Hoeksema:14} acquired from 2016 April~8 to April~19, with a time cadence of 12~minutes and a spatial resolution of 1\arcsec.  

We benefitted from high-resolution photospheric observations of the spectropolarimeter \citep[SP,][]{Lites:13} mounted on the Solar Optical Telescope \citep[SOT,][]{Tsuneta:08} aboard the \textit{Hinode} satellite \citep{Kosugi:07}. This instrument obtained a single raster scan from 02:21 to 03:24~UT on April~14, at the time of the passage at the central meridian of AR~12529 ($\mu = 0.96$). This SOT/SP scan along the \ion{Fe}{1} line pair at 630.15 and 630.25~nm has a pixel size along the slit of 0\farcs32, with a step size of 0\farcs32 and a step cadence of 3.8~s (Fast Mode). The region was scanned in 1000~steps, covering a field of view (FoV) of about $300 \arcsec \times 162\farcs3 $ that encompassed the whole AR~12529 (see Figure~\ref{fig:context}). 

To take into account the effect of the spatial point spread function (PSF) of the telescope, we performed a deconvolution of the original data which removes the induced stray light contamination. We used the regularization method proposed by \citet{PCA:13}, based on a principal component decomposition of the Stokes profiles, following the implementation of \citet{Quintero:16} for \textit{Hinode} SOT/SP data. In particular, we considered 15 principal components of eigenvectors for each Stokes profile, with 50 iterations for Stokes~\textit{I} and 25 iterations for the remaining Stokes parameters, using the spatial PSF for SOT/SP measurements with 0\farcs32 pixel size. As a consequence, the continuum contrast in a wide quiet-Sun region comprised in the FoV changed from 6.9\% in the original data to 12.1\% in the deconvolved data.

The deconvolved SOT/SP data were processed using the SIR \citep{SIR:92} inversion code to obtain the photospheric vector magnetic field. For this inversion, elemental abundances were taken from \citet{Asplund:09}. The SIR inversion yielded the temperature stratification in the range $-4.0 < \log \tau_{500} < 0$, where $\tau_{500}$ is the optical depth of the continuum at 500~nm. SIR also provided the line-of-sight (LOS) velocity, the micro-turbulent velocity, the magnetic field strength $B$, and the inclination and azimuth angles $\gamma$ and $\phi$ in the LOS reference frame.

In particular, we considered a sub-FoV of $130\arcsec \times 130\arcsec$ centered on the preceding sunspot of AR~12529, as shown in Figure~\ref{fig:context} (solide line). We divided this sub-FoV into three regions, identified by different thresholds of the normalized continuum intensity $I_c$ to account for the different physical conditions: quiet Sun ($I_c > 0.9$), penumbra ($0.5 < I_c < 0.9$), and umbra ($I_c < 0.5$). Following \citet{Murabito:16}, we used as an initial guess the temperature stratification of the Harvard-Smithsonian Reference Atmosphere \citep[HSRA;][]{HSRA} for the quiet-Sun model (also adopted for the plage of the AR, see \citealp{Guglielmino:18}), the values of $T$ provided by \citet{delToro:94} for the penumbral region, and the temperature stratification provided by \citet{Collados:94} for the umbra. In the latter, we adopted an initial value of 2000~G for the magnetic field strength, whereas an initial value of 1000~G was used for the penumbral region. Each inversion consisted of three iteration cycles. The temperature stratification was modified with up to three nodes (two for the first cycle, three for the others), whereas all other quantities were assumed to be constant with optical depth. Finally, the synthetic profiles were convolved with the spectral PSF at the focal plane of the instrument.

Some pixels belonging to the UF that showed peculiar profiles were also inverted considering a two-component atmosphere. In this case, for each component we used as initial guess of the temperature stratification both the above mentioned atmospheric models of the penumbral and of the umbral region. We imposed that both components should have the same temperature during the whole inversion process. Then, after several tries, we adopted an inversion scheme with three iteration cycles. The temperature stratification was modified with up to five nodes, whereas the magnetic field strength, the LOS velocity, and the inclination and azimuth angles were modified with one node in the first cycle, two nodes in the second one, and finally three nodes in the third one.

Although these SOT/SP measurements were taken very close to disc center, nevertheless we transformed the inclination and azimuth angles $\gamma$ and $\phi$ from the LOS reference frame into the Local Reference Frame (LRF). To obtain this transformation, first we resolved the $180^{\circ}$ azimuth ambiguity. There exist several methods to solve the $180^{\circ}$ azimuth ambiguity (see the review of \citealp{Metcalf:06}). We used the Acute Angle Method, i.e., we compare the inverted magnetic field to a potential extrapolated field. The azimuth is such that the extrapolated and inverted fields make an acute angle between them \citep{Metcalf:06}. Then, the $\gamma_{\mathrm{LOS}}$ and the disambiguated $\phi_{\mathrm{LOS}}$ angles were converted into the LRF by applying the transformations described by \citet{Gary:90}.

To calibrate Doppler velocity values, we assumed that umbra (i.e., pixels with normalized continuum intensity $I_c < 0.5$) is on average at rest \citep[e.g.][]{Rimmele:94}. 

\begin{figure*}[t]
	\centering
	\includegraphics[trim=0 190 0 60, clip, scale=0.6075]{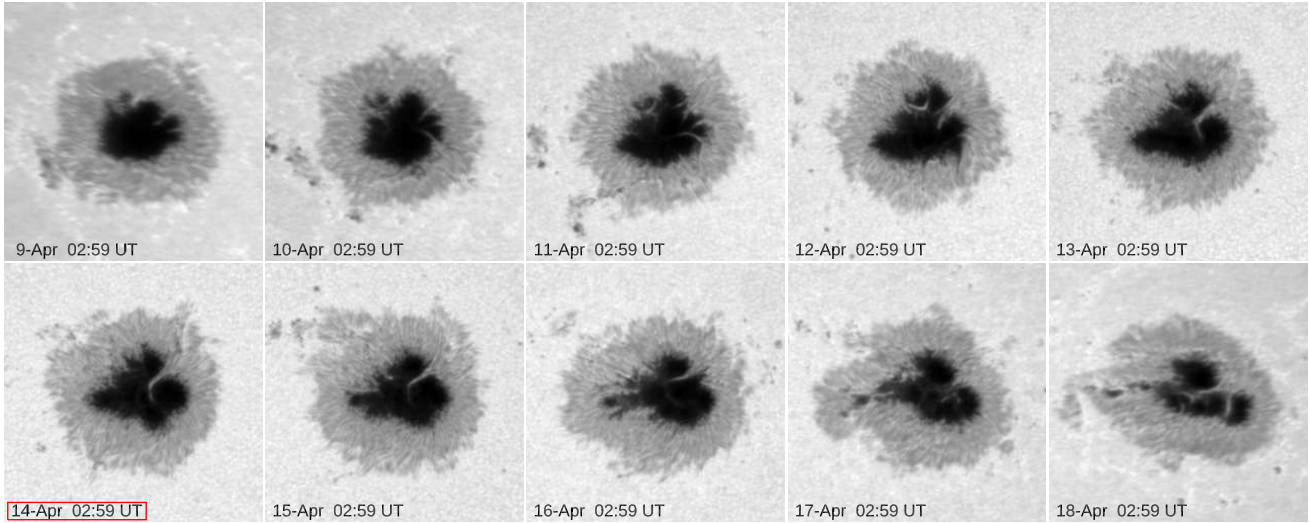}
	\caption{Evolution of the main sunspot of AR NOAA~12529 between 2016 April~9 and April~18 as seen in the continuum filtergrams taken by SDO/HMI in the \ion{Fe}{1} 6173~\AA{} line, with a cadence of one day. The FoV of these images is the same framed with a box in Figure~\ref{fig:context}. The image taken on April~14, with the hour framed by a red box, is the closest to the \textit{Hinode} SOT/SP observations at the central meridian. 
	\label{fig:sdo}}
\end{figure*}

The \textit{IRIS} satellite acquired two data sets in the ultraviolet (UV) relevant to the preceding sunspot of AR~12529 at the time of its central meridian passage. These consist of single, very large, dense 400-step rasters (observing sequence OBS3610108078), with simultaneous slit-jaw images (SJIs) composed of 1330, 1400, 2796, and 2832~\AA{} filtergrams. Both rasters were acquired on April~14: the first was made between 04:27 and 05:28~UT, the second between 05:39 and 06:40~UT. The sequence had a 0\farcs33 step size, a 9~s step cadence, and a 8~s exposure time. The pixel size is 0\farcs35, and the FoV of each scan is $140\farcs5 \times 182\farcs3$. SJIs have a cadence of about 37~s for consecutive frames in each passband, covering a total FoV of $308\farcs8 \times 182\farcs3$. The usable FoV of each frame of SJIs is about $168 \arcsec \times 178\arcsec$. 
The \textit{IRIS} data were downloaded as level~2 products, already reduced by the instrument team. The version of the calibration processing IDL \textit{Solarsoft} routine (\textsc{iris\textunderscore{}prep}) applied to the data was 1.56.

Finally, we also used images acquired in the extreme ultraviolet (EUV) by the Atmospheric Imaging Assembly \citep[AIA;][]{Lemen:12} aboard the \textit{SDO} satellite at 304 \AA{}, with a pixel size of about 0\farcs6, simultaneously to the \textit{IRIS} observations.

\section{Results}

Figure~\ref{fig:sdo} presents the evolution of the large preceding sunspot of AR~12529 during the entire passage across the solar disc in April 2016. The giant sunspot appeared already formed at the East limb on April~8, and was still visible when it rotated behind the West limb on April~19, as mentioned in Paper~I. It occupied an area of $\approx 2000 \,\mathrm{Mm}^2$, being about fifteen times as large as the area of the Earth's disc. 

The initial frames of the sequence in Figure~\ref{fig:sdo} display a rather roundish sunspot umbra, into which some penumbral filaments occasionally intrude, specially to the north-eastern edge of the spot. They had a maximum lifetime of about one day. 
Conversely, the penumbra experienced repeated fragmentation episodes since the beginning of the observations: for instance, in the south-eastern part (e.g., in the intervals [9-Apr 02:59 UT -- 10-Apr 02:59 UT] and [10-Apr 02:59 UT -- 11-Apr 14:59 UT]) and in the north-eastern part (in the interval [14-Apr 02:59 UT -- 15-Apr 14:59 UT]). Starting from April~16 the sunspot, at that time appearing rather elongated along the East-West direction, underwent a severe decay phase: it fragmented and LBs that split the umbra into several smaller and smaller umbral cores appeared, as observed until the end of the observations.

The passage of AR~12529 at the central meridian occurred between April~13 and~14. The frame closest in time to the passage is indicated in Figure~\ref{fig:sdo} by a red box (second row, first column). At that time, the giant spot looked peculiarly heart-shaped, owing to the presence of an intruding structure to the North-West edge of the sunspot umbra (see also Figure~\ref{fig:context}, acquired simultaneously). This feature is the UF analyzed in Paper~I. Figure~\ref{fig:sdo} points out that the lifetime of the UF is about four days. It was detectable on April~12 at 02:59~UT and was lastly seen on April~16 at 02:59~UT.

\begin{figure*}[t]
	\centering
	\includegraphics[trim=0 15 180 30, clip, scale=0.8]{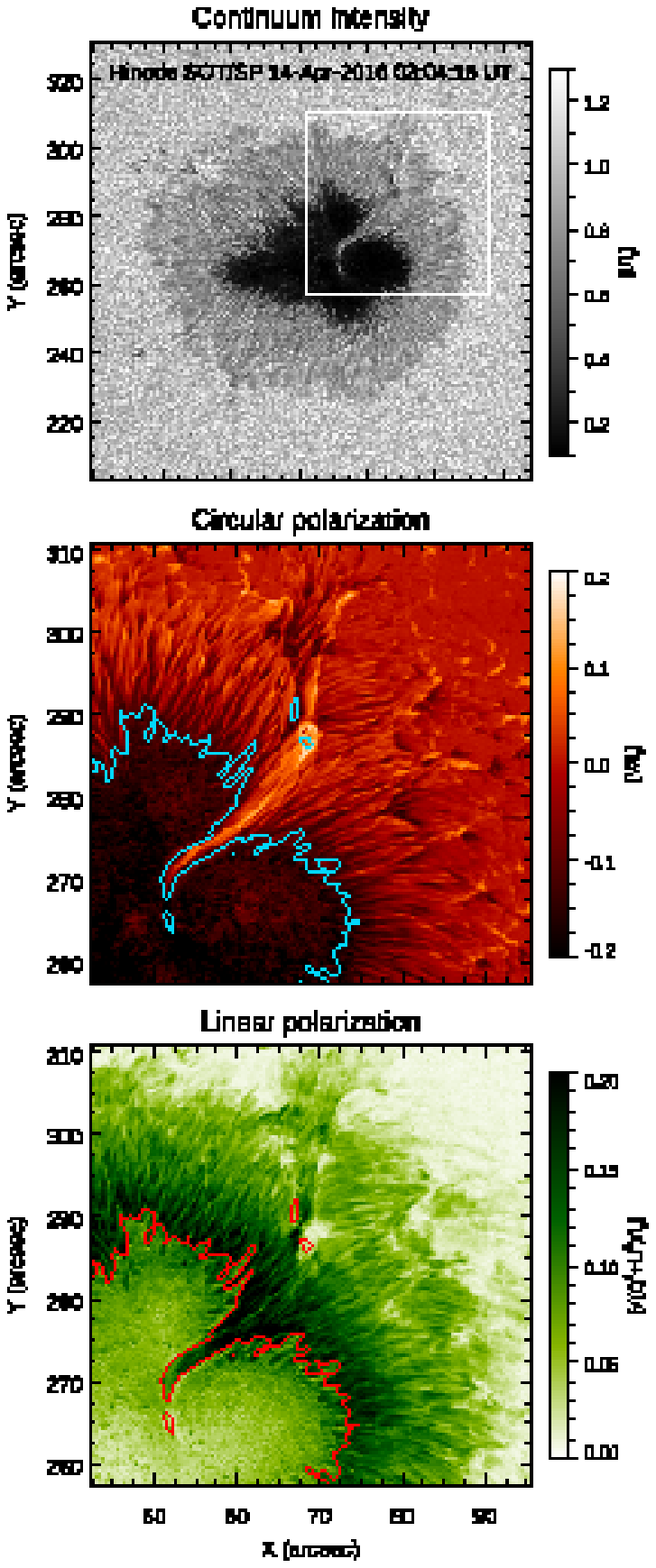}%
	\includegraphics[trim=15 15 0 30, clip, scale=0.8]{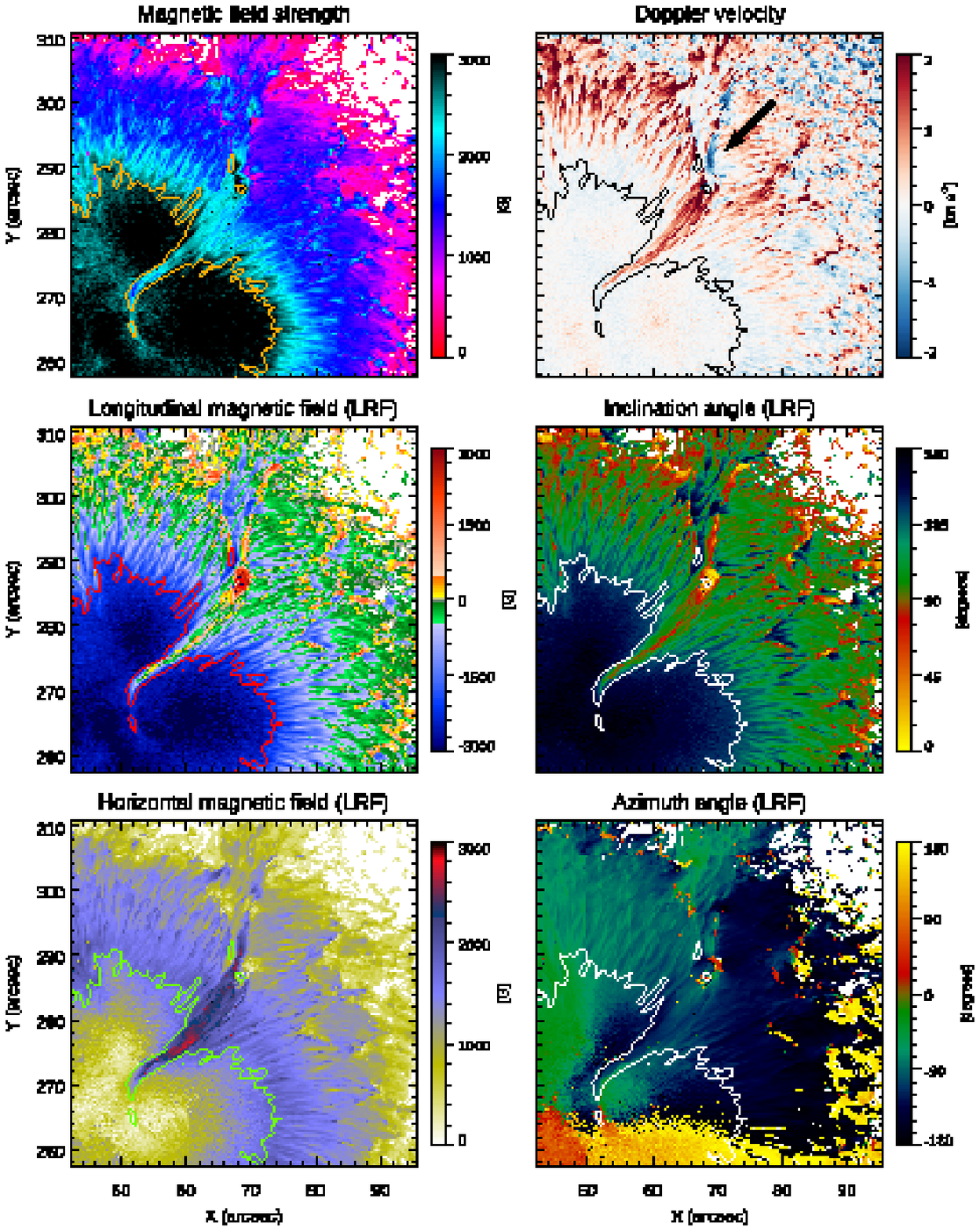}
	\caption{Left column: Map of the continuum intensity for the sub-FoV indicated with a black box in Figure~\ref{fig:context} (top panel). The white box frames the RoI studied in detail in the other panels. Maps of the circular polarization signal (middle panel) and linear polarization signal (bottom panel), for the RoI indicated with a white box in the continuum map. Central and right columns: Maps of the physical quantities retrieved by the SIR inversion for the RoI indicated with a white box in the top-left panel: magnetic field strength, longitudinal and horizontal component of the vector magnetic field, Doppler velocity, inclination and azimuth angles of the vector magnetic field. The contours represent the continuum intensity at the umbra boundary, $I_{c}=0.5$. The arrow in the top-right panel points to the penumbral gap. \label{fig:sot}}
\end{figure*}

Figure~\ref{fig:sot} (left column) shows the maps deduced from spectropolarimetric measurements acquired by \textit{Hinode} SOT/SP at the time of the passage at the central meridian of AR~12529. The UF is clearly seen within the umbra of the preceding giant sunspot in the continuum intensity map, at about solar heliocentric position $\mathrm{X}=[50\arcsec, 60\arcsec] \times \mathrm{Y}=[270\arcsec, 280\arcsec]$. The leading edge of the UF seems to correspond with a penumbral gap, while its trailing edge ends quite abruptly well inside the umbra. Even with the \textit{Hinode}/SOT resolution, about $0\farcs3$ at 630~nm, no granular pattern is found in the filamentary structure. The map of circular polarization (middle panel, left column), relevant to the region of interest (RoI) framed with a white box in the continuum map, reveals that the signal in a wide part of the UF has opposite sign with respect to that of the hosting umbra. Interestingly, the linear polarization map for the same RoI (bottom panel, left column) indicates a strong signal in the area of the UF, much larger than in the surrounding penumbral filaments. It is worth recalling that the use of the PCA deconvolution method ensures that these polarization signals inside the UF are fully reliable, since they cannot be due to contamination with the signals coming from the surrounding magnetic elements.

The inversion of the SOT/SP spectropolarimetric measurements with the SIR code provides additional information about the configuration and the physical condition of the plasma in the UF. In Figure~\ref{fig:sot} (middle and right columns), we display zoomed maps of the physical parameters retrieved by the inversion, for the RoI framed with a white box in the continuum map in Figure~\ref{fig:sot} (top-left panel). From top to bottom, from left to right, they are magnetic field strength, Doppler velocity, longitudinal component of the magnetic field, inclination angle, horizontal component of the magnetic field, and azimuth angle, being the latter quantities already transformed into the LFR. These maps point out peculiar characteristics of the UF, described as follows.
\begin{itemize}[noitemsep]
	\item The UF has a magnetic field strength generally larger than 2000~G, with small portion of it with a minimum value of about 1500~G.
	\item The flow pattern along the structure exhibits a redshift along the part of the structure inside the umbra that becomes stronger toward the penumbra. With an abrupt discontinuity, a blueshift is observed at the egde of the sunspot, where there is a small gap in the main penumbral filaments.
	\item The longitudinal field map as well as the inclination angle map indicate the presence of the opposite magnetic polarity, with respect to the sunspot umbra, along the interior of the UF. 
	\item The UF harbors strong horizontal fields, with some patches having a horizontal field even larger than 2500~G. These fields are substantially stronger than the horizontal field in the surrounding penumbral filaments, which reaches values up to 1500~G.
	\item The azimuth angle in the UF region is not as coherent as in the rest of the sunspot, possibly indicating a kind of divergence of the magnetic field lines: apparently, the lines separate from the axis of the filament as we move toward the umbra. Moreover, it exhibits an axial symmetry along the axis of the UF. Note that the azimuth points from $-180^{\circ}$ to $-90^{\circ}$, from the east side toward the west side and from north to south of the UF, with a rather abrupt change. 
\end{itemize}

\begin{figure*}[t]
	\centering
	\includegraphics[trim=15 25 60 335, clip, scale=0.4175]{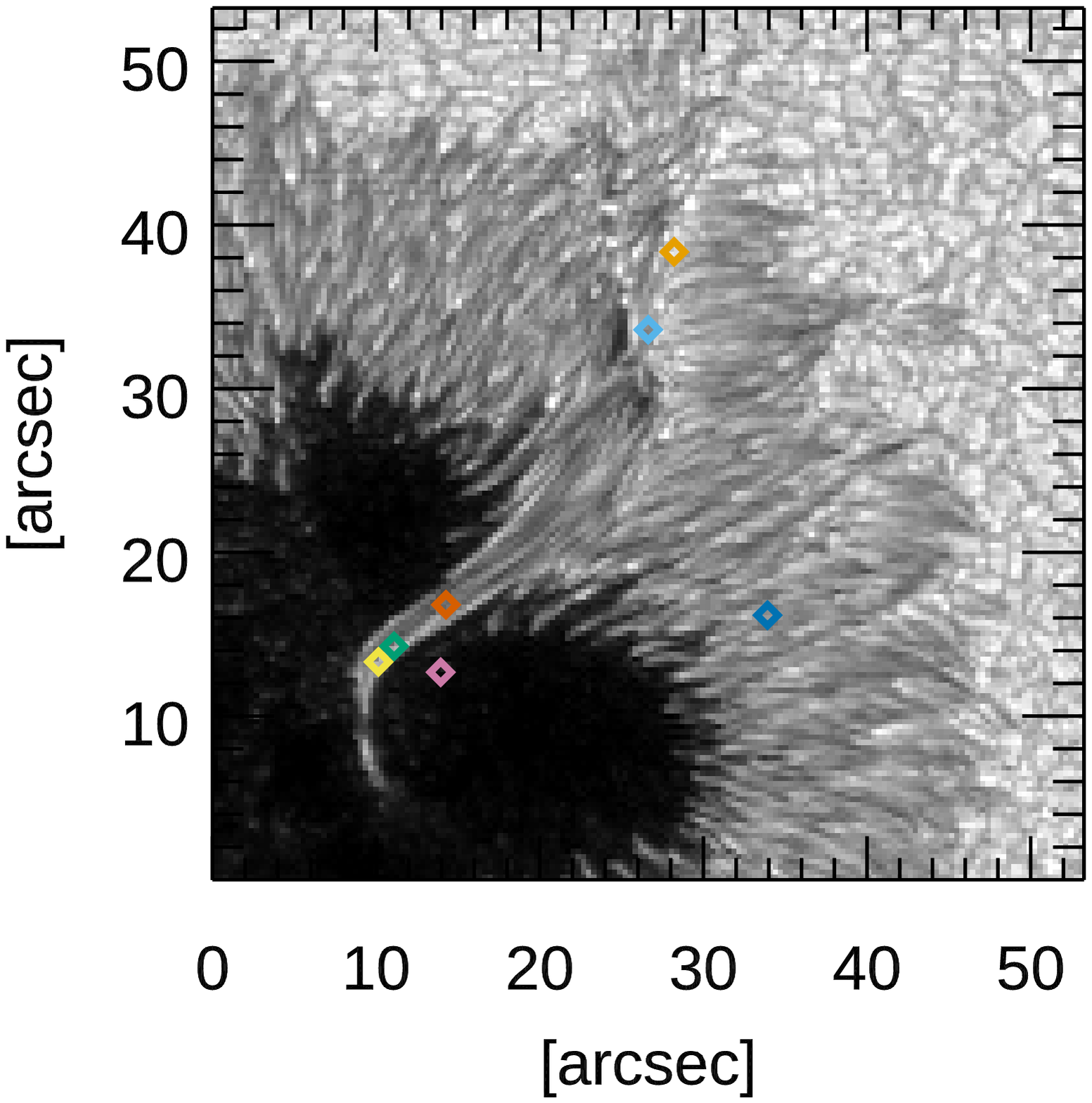}%
	\includegraphics[trim=0 17 54 167, clip, scale=0.6]{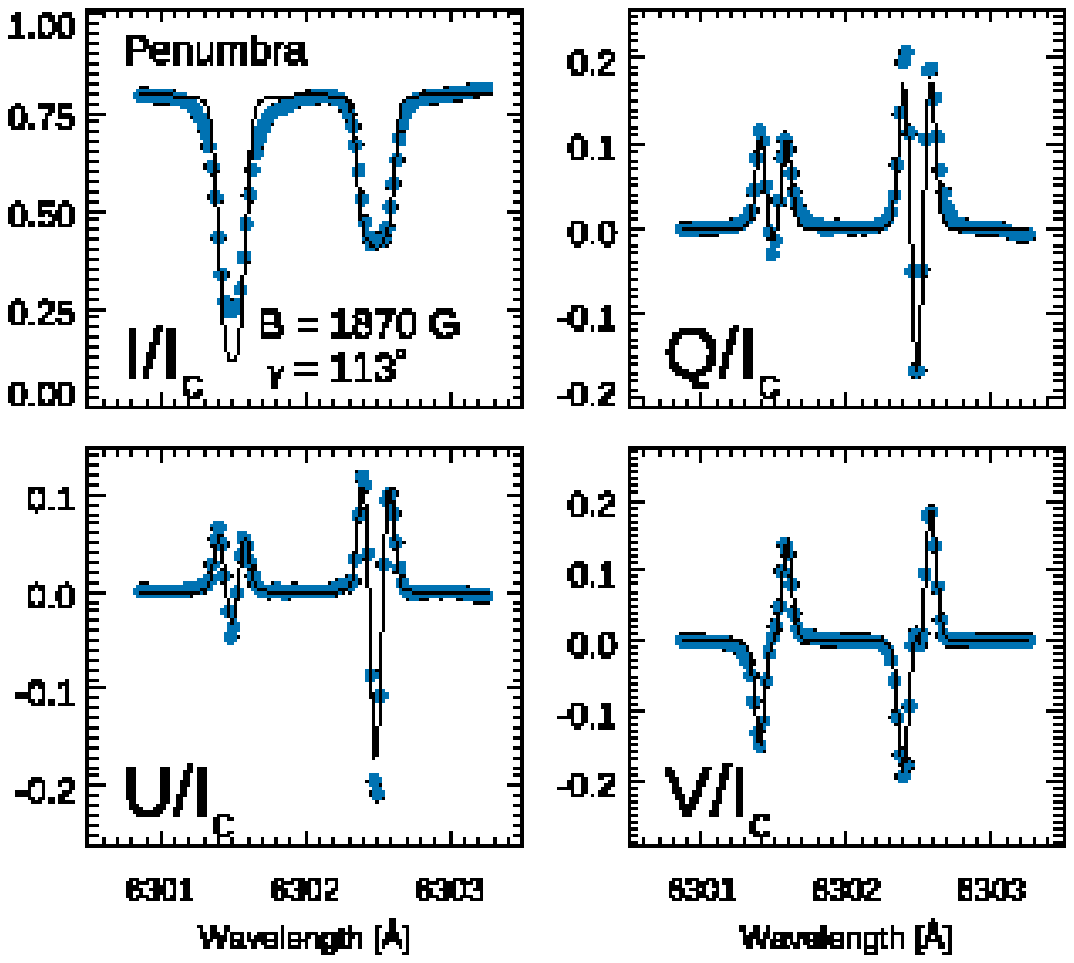}		
	\includegraphics[trim=5 37 32 232, clip, scale=0.6]{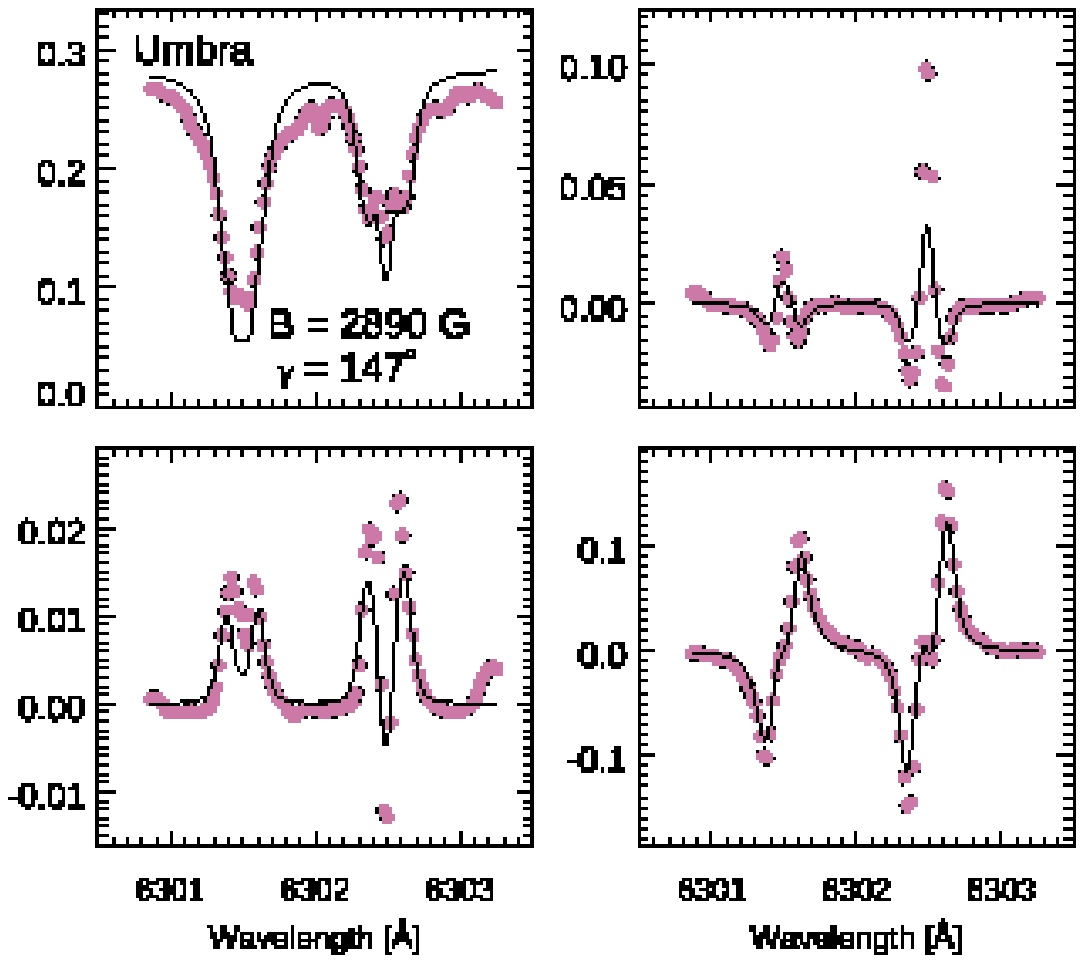}%
	\includegraphics[trim=10 37 54 232, clip, scale=0.6]{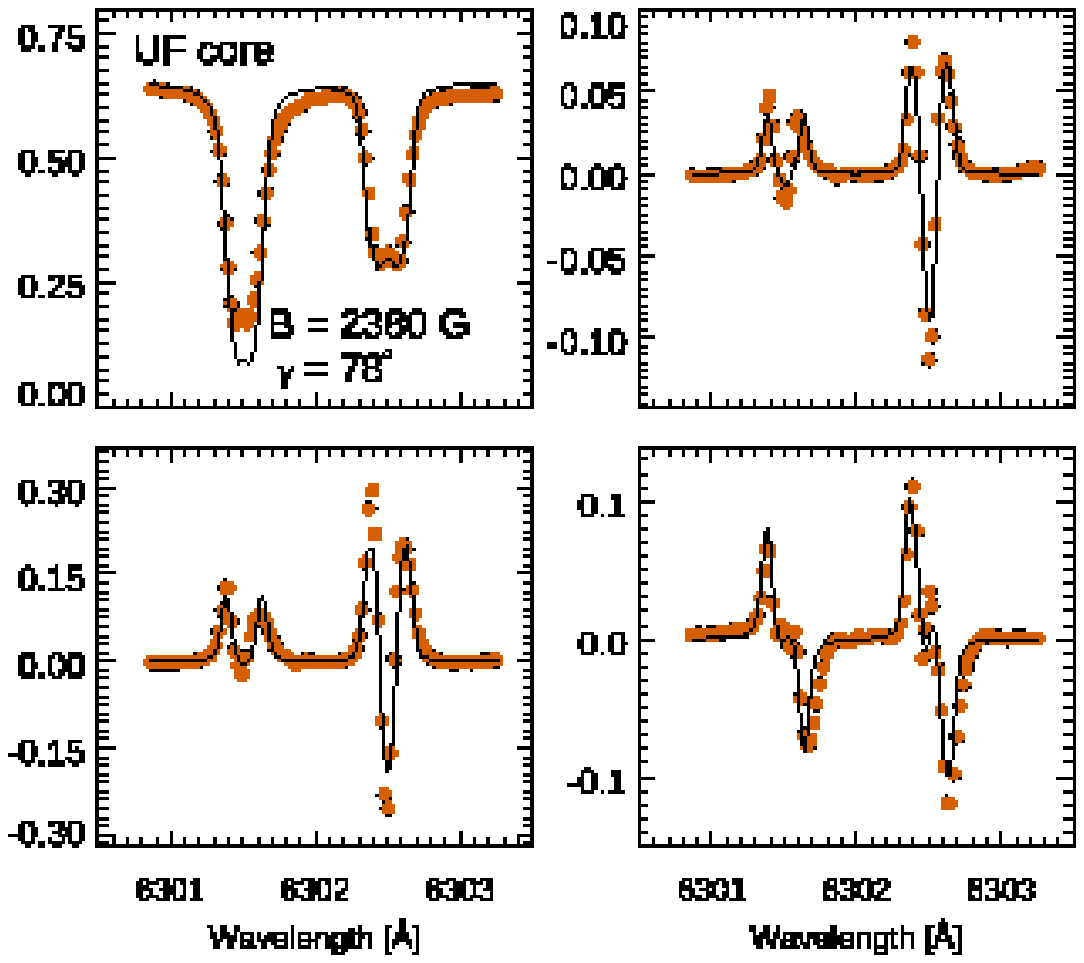}
	\includegraphics[trim=5 37 32 232, clip, scale=0.6]{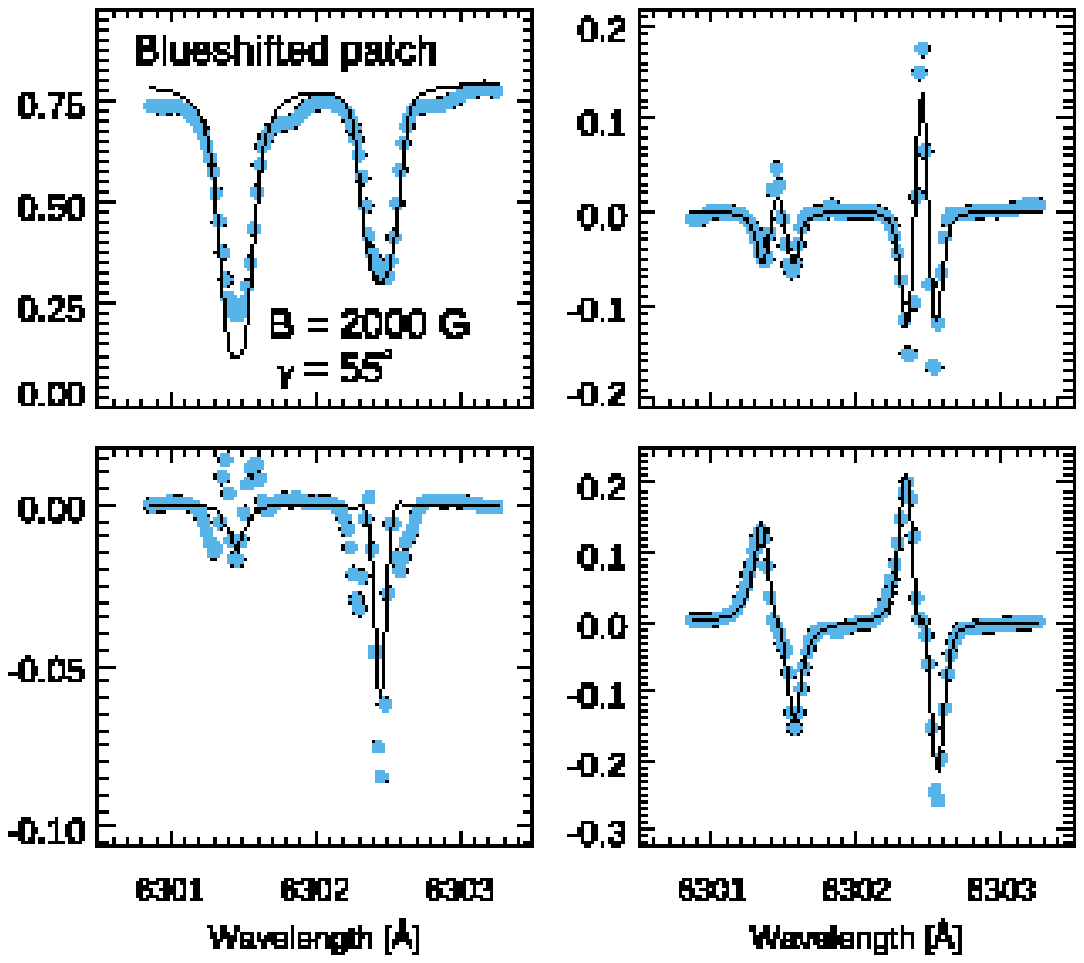}%
	\includegraphics[trim=10 37 54 232, clip, scale=0.6]{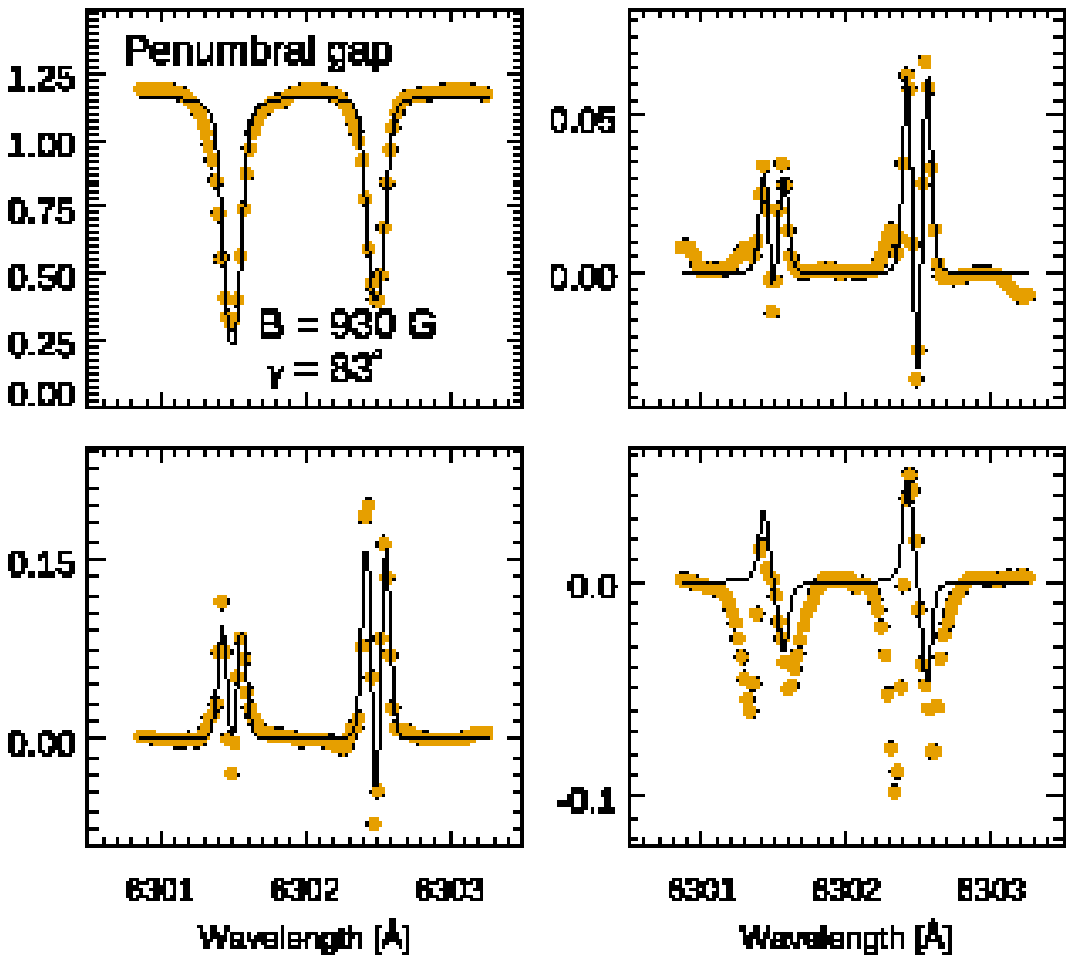}
	\caption{Top-left panel: Reconstructed continuum map for the RoI indicated with a white box in Figure~\ref{fig:sot}. The colored diamonds indicate the pixels whose Stokes profiles are shown in this same Figure and in Figure~\ref{fig:peculiar}. Other panels: Stokes profiles measured along the \ion{Fe}{1} 630.25~nm pair (filled circles) and their fits (solid lines), as provided by the SIR inversion, color-coded according to the pixels pinpointed in the continuum map (top-left panel). \label{fig:profiles}}
\end{figure*}

\begin{figure*}
	\centering
	\includegraphics[trim=10 75 0 335, clip, scale=0.435]{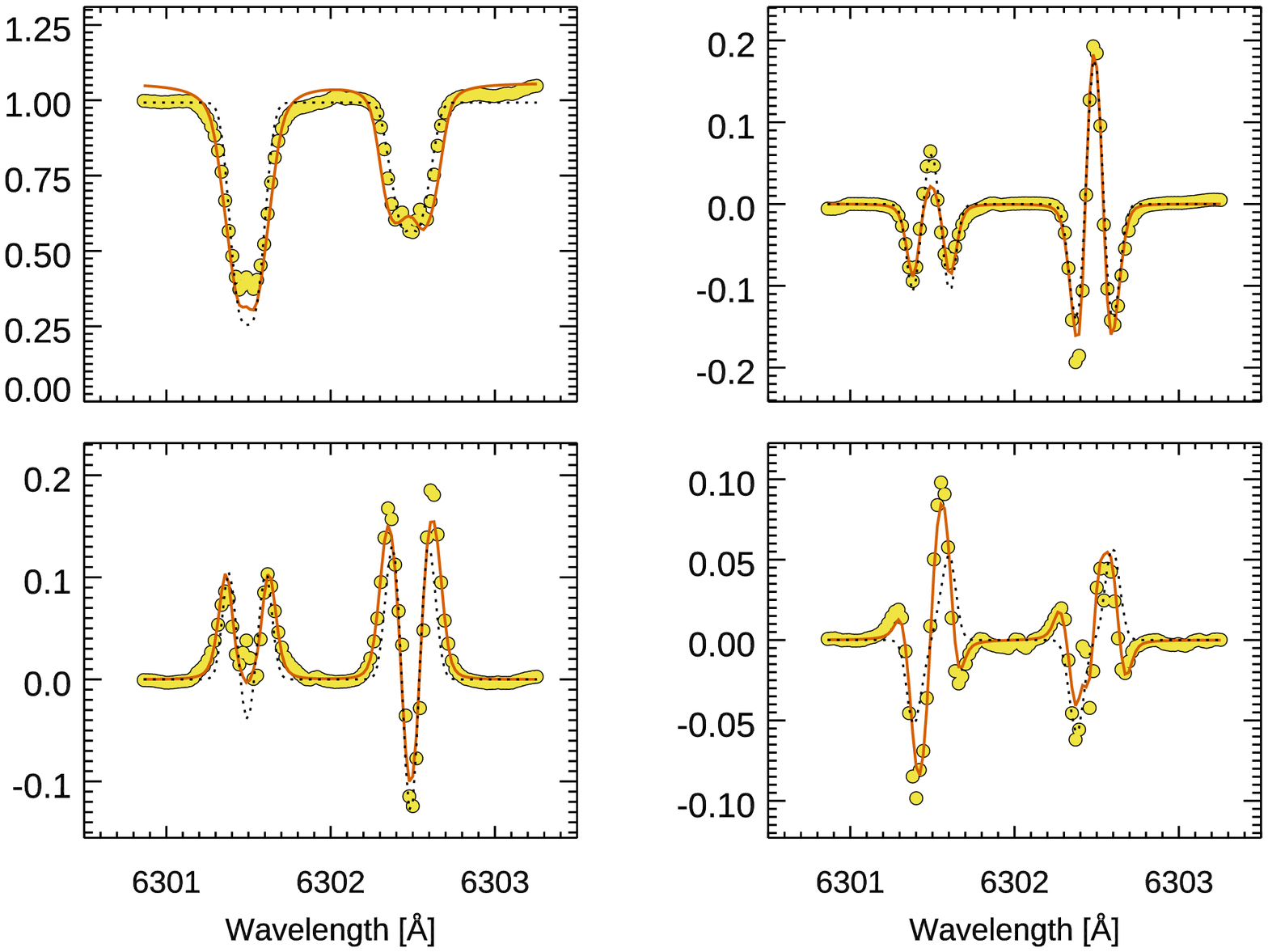}%
	\includegraphics[trim= 5 75 5 335, clip, scale=0.435]{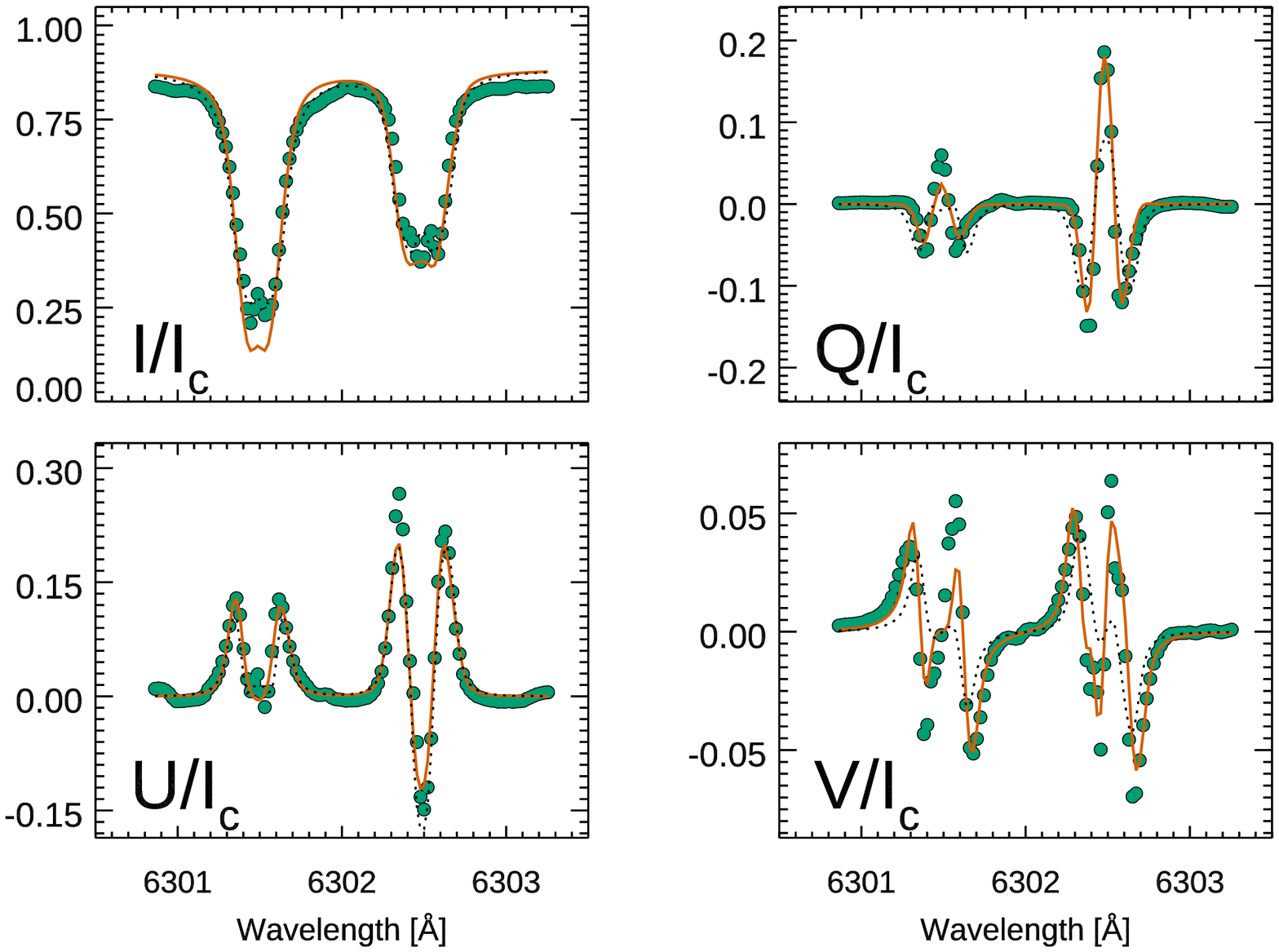}	
	\includegraphics[trim=10 65 0 335, clip, scale=0.435]{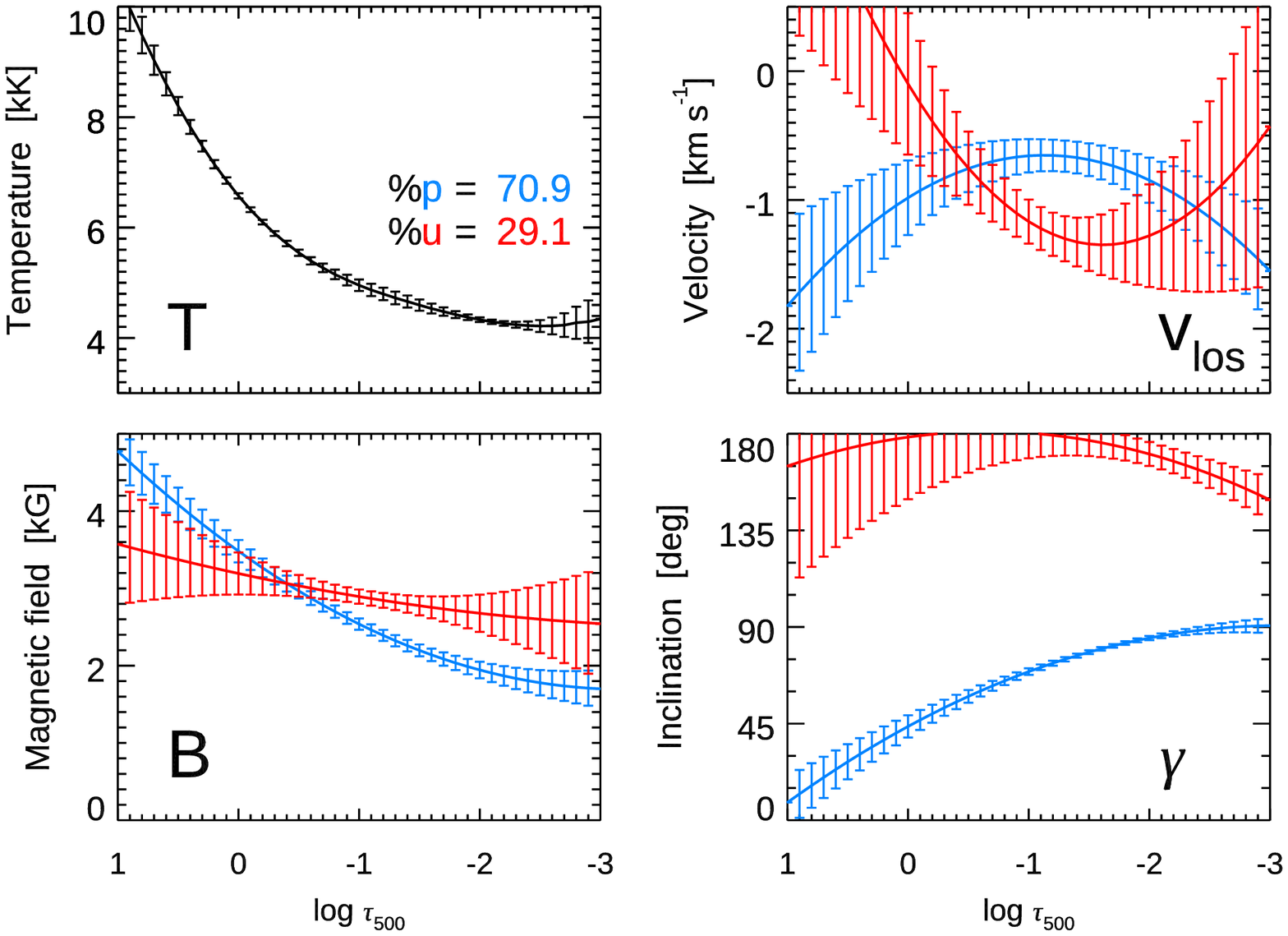}%
	\includegraphics[trim= 5 65 5 335, clip, scale=0.435]{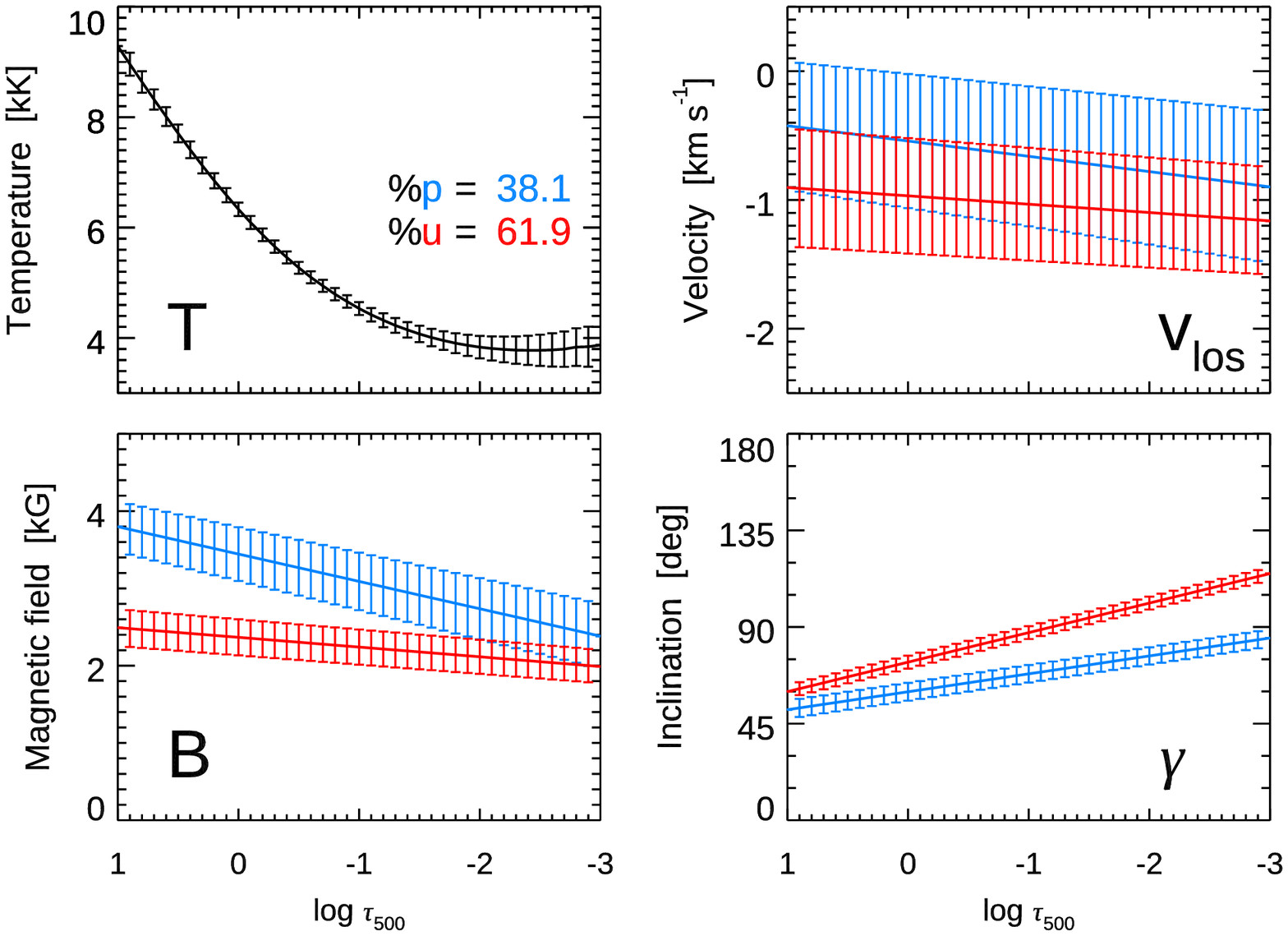} 
	\caption{Top panels: Observed profiles and fits obtained with the SIR double-component (solid red line) and single-component (dashed black line) inversions for those pixels indicated with the yellow and green diamonds in the continuum map shown in Figure~\ref{fig:profiles}. Bottom panels: the atmospheric stratification of (clockwise, from the top-left panel) temperature, Doppler velocity, magnetic field strength, and inclination angle of the vector magnetic field as deduced from the SIR double-component inversions. The percentage of \textit{p} and \textit{u} indicate the relative filling factor of the penumbral and umbral components, which are indicated in blue and red colors, respectively. \label{fig:peculiar}}
\end{figure*}

It is also worthy detailing some properties of the penumbral gap observed in the continuum at about solar heliocentric position $\mathrm{X}=70\arcsec , \, \mathrm{Y}=[285\arcsec, 305\arcsec]$ (see the arrow in Figure~\ref{fig:sot}, top-right panel), where a Doppler velocity discontinuity occurs at the edge of the UF. As a matter of fact, an elongated patch presents blueshift. This blue-shifted patch has positive polarity, opposite to that of the hosting sunspot, with field strength of $1500 - 2000$~G. The magnetic field strength reaches values up to $3000$~G in the region comprised between the small dark knots visible in the continuum map in the inner part of the penumbral gap, closer to the umbra. In the same area, the vector magnetic field exhibits a polarity discontinuity and has a large horizontal component, about $2500$~G (Figure~\ref{fig:sot}, central column, middle and bottom panels).

In Figure~\ref{fig:profiles} we display the observed Stokes profiles (filled circles) for some representative locations of the RoI analyzed in Figure~\ref{fig:sot}, as indicated by diamonds in the continuum map (top-left panel). We also plot the fits to the data obtained with the single-component SIR inversion (solid lines). The blue profile refers to a pixel of the normal penumbra surrounding the spot that has a quite inclined magnetic field ($\gamma = 113^{\circ}$) with a strength lower than $1900$~G. The magenta pixel represents the umbral region, with strong field up to $\approx 2900$~G and nearly vertical ($\gamma= 147^{\circ}$). Finally, the red pixel is relevant to the UF core. The Stokes profiles in the UF core are almost similar to those of the normal penumbra, but with the magnetic field clearly characterized by opposite polarity with respect to that of the sunspot. (Note the reversed Stokes~\textit{V} profile with respect to that of the blue and magenta pixels). Interestingly, in the UF the magnetic field is rather horizontal, with an inclination angle of $78^{\circ}$, and has remarkably stronger intensity ($\approx 2400$~G) than in the regular penumbra. Conversely, the magnetic field strength is about $500$~G lower than in the umbra. We also show some profiles in the penumbral gap region. The light blue profile is relevant to the blue-shifted patch seen in the Doppler velocity map in Figure~\ref{fig:sot}. It exhibits a strong magnetic field, with intensity of $2000$~G, and opposite polarity with respect to that of the hosting umbra. Slightly to the North, in the outer part of the penumbral gap (orange profile), we observe a rather horizontal field, with strength of about $1000$~G. In this point, a multi-lobed Stokes~\textit{V} profile is observed, indicating mixed polarities. While Stokes~\textit{Q} and~\textit{U} are fitted quite well by using the single-component inversion scheme, the fit of Stokes~\textit{V} has clearly a poor quality. The observed Stokes~\textit{V} profile is compatible with the coexistence in the same pixel of two different atmospheres, characterized by opposite magnetic polarities and different LOS velocities (possibly one being blue-shifted and the other red-shifted). Therefore, a single-component inversion is unable to fit such a profile. This makes clear the need of a more accurate two-component inversion for certain pixels.

Figure~\ref{fig:peculiar} (top panels) presents the Stokes profiles for the pixels indicated with a yellow and a green diamonds in Figure~\ref{fig:profiles}, showing together their fits obtained with both the single-component (dotted black line) and double-component (solid red line) SIR inversions. 
These pixels are located near the periphery of the UF, close to the umbra. The different quality of the fits seen in Figure~\ref{fig:peculiar} (top panels) uncovers once again the need of two atmospheric components in each pixel in order to get a proper modelling of the measured Stokes profiles. 

The retrieved physical parameters for the profiles relevant to the yellow pixel (Figure~\ref{fig:peculiar}, bottom-left panels) indicate that at $\log \tau_{500} = -1.5$ the magnetic field of the red component is $\approx 3000$~G, stronger than the field of the blue component (about $2000$~G), the latter being more horizontal (inclination angle $\approx 85^{\circ}$) than the former (almost $180^{\circ}$). This provides the classical view for the penumbra: a prevailing mostly horizontal structure of flux tubes (intra-spines) embedded in a more vertical background field (spines), with stronger magnetic field intensity than in the flux tubes \citep[e.g.,][see the Discussion]{Bellot:04,Tiwari:15}. In this sense, the red component that has the same polarity of the umbra, which we can refer to as ``umbral'' component, matches with the classical spines component and the blue component with the intra-spines, being referred to as ``penumbral'' or ``flux-tube'' component. In the second case, for the green pixel, the flux-tube component has a strong magnetic field ($\approx 3000$~G) and it is slightly horizontal ($\gtrsim 70^{\circ}$), whereas the umbral component is weaker and even more horizontal than the penumbral component. Note that, in both examples, the flux-tube component has positive magnetic polarity. On the other hand, the information about the relative difference deduced in the Doppler velocity between the two components is inconclusive, as they appear to have the same Doppler velocity within the error bars. 

\begin{figure}[t]
	\centering
	\includegraphics[trim=15 15 225 0, clip, scale=0.4]{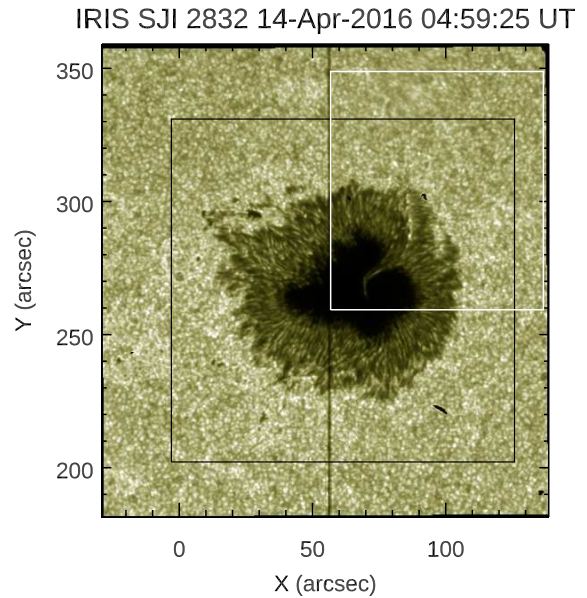}
	\caption{Map deduced from \textit{IRIS} observations of the AR~12529 acquired at 04:59 UT on 2016 April 14, in the 2832~\AA{} band (photosphere). The rectangular white box frames the area imaged in Figure~\ref{fig:iris}. For comparison, the black box points to the \textit{Hinode} SOT/SP sub-FoV indicated with a box in Figure~\ref{fig:context} and analyzed in Figure~\ref{fig:sot}. \label{fig:irisfov}}
\end{figure}

\begin{figure}[b]
	\centering
	\includegraphics[trim=10 0 115 10, clip, scale=0.8]{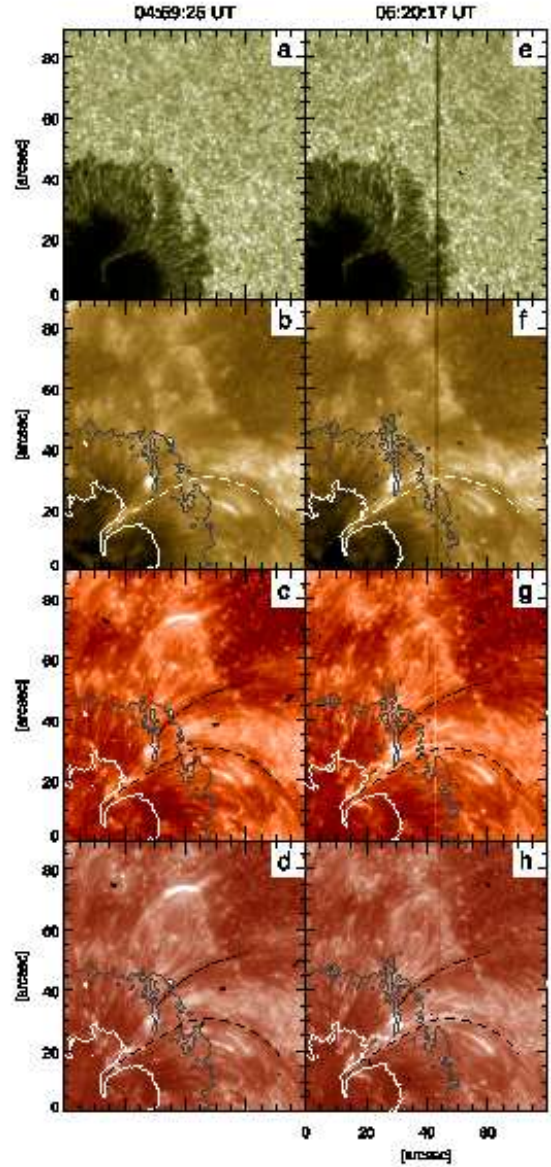}
	\caption{\textit{IRIS} SJ images in different passbands: 2832~\AA{} panels (a)/(e); 2796~\AA, panels (b)/(f); 1330~\AA, panels (c)/(g); and 1400~\AA, panels (d)/(h), for two different times during the first \textit{IRIS} observing sequence on April~14. For comparison, the contours indicate the umbra-penumbra boundary (white) and penumbra-moat boundary (grey) in the 2832~\AA{} passband. The dashed curved lines indicate the direction along which elongated brightenings are observed (see the main text), corresponding to filament 1 in Figure~\ref{fig:aia}. The dash-dotted lines indicate filament 2 in Figure~\ref{fig:aia}. \label{fig:iris}}
\end{figure}

In order to study the counterpart of the UF in the upper atmospheric layers, we considered the high-resolution filtergrams relevant to the leading spot of AR~12529 acquired by \textit{IRIS}, almost simultaneous to the \textit{Hinode} SOT/SP measurements. Figure~\ref{fig:irisfov} displays the FoV of \textit{IRIS} SJ observations in the 2832~\AA{} band. For comparison, we indicate with a black box the sub-FoV of \textit{SDO}/HMI and \textit{Hinode} SOT/SP observations previously analyzed. The white box frames the \textit{IRIS} sub-FoV where the UF is located, which is imaged in Figure~\ref{fig:iris}. 

The sequence of images relevant to the four UV bands of \textit{IRIS} SJIs (Figure~\ref{fig:iris}) illustrates the chromospheric and transition region counterpart of the UF and its evolution. Panels (a) and (e) reveal slightly enhanced emission in the 2832~\AA{} passband in the region cospatial to the penumbral gap near the leading edge of the UF, around X=30\arcsec, Y=[25\arcsec, 35\arcsec]. A compact burst, with increasing intensity, is seen near the base of the penumbral gap (X=30\arcsec, Y=30\arcsec) in the \ion{Mg}{2} band (2796~\AA, panels (b) and (f)). Moreover, elongated brightenings are observed along a curved direction (dashed line) that departs from the trailing edge of the UF inside the umbra. See, for instance, the elongated bright feature centered at X=55\arcsec, Y=20\arcsec{} in panel (b). Correspondingly, brightness enhancements are also observed in the \ion{C}{2} (1330~\AA, panels (c) and (g)) and \ion{Si}{4} (1400~\AA, panels (d) and (h)) bands. These \ion{C}{2} and \ion{Si}{4} SJIs look similar to each other, exhibiting a bright knot, cospatial to that seen in the \ion{Mg}{2} band although weaker in intensity, and elongated brightenings along the same curved direction as noticed before (dashed lines). However, the morphology of these brightenings appears to be rather complex, especially in the surroundings of the UF, along which bright fibrils seem twisting. The comparison of panels (c) and (d) to panels (g) and (h), respectively, indicates that the elongated brightenings change in intensity with time. A visual inspection of the intermediate filtergrams suggests that some of these brightenings may be flowing along the curved direction. Finally, we note also that darker structures, observed in panels (c)--(h), follow the same paths as the elongated brigthenings.

\section{Discussion and conclusions}

In this work, we analyzed the properties of the UF intruding the giant leading sunspot of AR~12529. This elongated structure had resemblance to an unsegmented LB, but exhibited completely different physical characteristics. Aiming at providing further observational evidence to the findings reported in Paper~I, we used high-resolution observations acquired in the photosphere by the \textit{Hinode} SOT/SP and in the chromosphere and transition region by the \textit{IRIS} spacecraft at the time of the passage at the central meridian of AR~12529. 

With regard to the photospheric measurements, we took advantage of a spatial deconvolution technique that removes the stray light contamination induced by the spatial PSF of the \textit{Hinode}/SOT telescope, implemented by \citet{Quintero:16}. This method is able to reduce the bias due to the stray light, which causes a lack of contrast in continuum images and a reduction of the amplitude of the magnetic Stokes profiles \citep{PCA:13,Quintero:15}. Thus, the polarimetric signals here studied are genuinely due to the magnetic fields that were present in each resolution element of the instrument. Furthermore, the contrast is enhanced, resulting in sharper continuum images.

The new information conveyed by this analysis complements the previous results and allows concluding that the interpretation of this UF as a kind of LB must be definitely discarded. The absence of any granular pattern in the UF, even at the diffraction-limited resolution of \textit{Hinode}/SOT, allows us to exclude that this structure is a segmented (or granular) LB. Moreover, the detection of a very large horizontal component of the magnetic field along the UF unquestionably rules out the possibility that such a structure may be a FLB. In fact, any kind of LB would be characterized by a field-free configuration or, at least, by a region of weak magnetic fields. 

In this respect, the strong horizontal component with filamentary appearance is reminiscent of the structure of regular penumbral filaments, although the azimuth angle and polarity of its magnetic field differs significantly from the surrounding penumbral filament. Indeed, we can compare the properties observed in the UF to the magnetohydrostatic penumbral model proposed by \citet{Borrero:10}. In that model, they considered the presence of the Evershed flow along the axis of the filament and a convective velocity perpendicular to it, finding that a field of about 1000~G was necessary to reproduce some observational properties, such as the net circular polarization. This model predicts that penumbral filaments, being represented by flux tubes, should exhibit downflows at both the edges of the central axis as a consequence of the convective motions. Moreover, the azimuth angle of the vector magnetic field should be perpendicular to the axis of the filament, pointing to opposite directions at each lane. Indeed, both these characteristics, that is, redshifts (i.e., downflows) along the lanes parallel to the main axis of the UF and azimuth perpendicular to it pointing to diverging directions in the lanes, are observed in the \textit{Hinode} SOT/SP measurements here analyzed. 

Therefore, at a first glance, it may seem that the UF here analyzed is nothing but a giant filament that developed and intruded in the sunspot. For the sake of completeness, intruding filaments are sometimes observed in numerical simulations of sunspots. For instances, they occur in the three-dimensional sunspot model described by \citet{Rempel:15} and analyzed by \citet{Siu-Tapia:18} (see their Figure~2). However, they are transient and do not exhibit the properties here detailed for the UF of AR~12529. One can note that similar transient features are observed in the leading sunspot of AR~12529 during its passage across the solar disc. These are noticed in the photospheric continuum (see Figure~\ref{fig:sdo}). 
Moreover, if the UF were due to the intrinsic field of the sunspots, the model proposed by \citet{Borrero:10} would not be able to account in a proper manner either for the opposite polarity seen in a large portion of the observed UF. As a matter of fact, the profiles studied in representative locations of the UF indicate that, on the one hand, the polarity of a large portion in the core of the feature is opposite to that of the hosting umbra. On the other hand, the pixels relative to the periphery of the UF need to be modelled by using a multi-component atmosphere, in which the flux-tube component has often a strong horizontal field, up to $\approx 2500$~G, with polarity opposite to that of sunspot. According to this model, the only possibility for the existence of the opposite polarity with respect to the hosting sunspot would be overturning convection, able to wrap around the magnetic field lines in the downflows of the lanes. Nonetheless, this would not occur in the center of the filament but toward the outer part, as confirmed by observations \citep[e.g.][]{Joshi:11}.

\begin{figure}[t]
	\centering
	\includegraphics[trim=25 15 195 160, clip, scale=0.425]{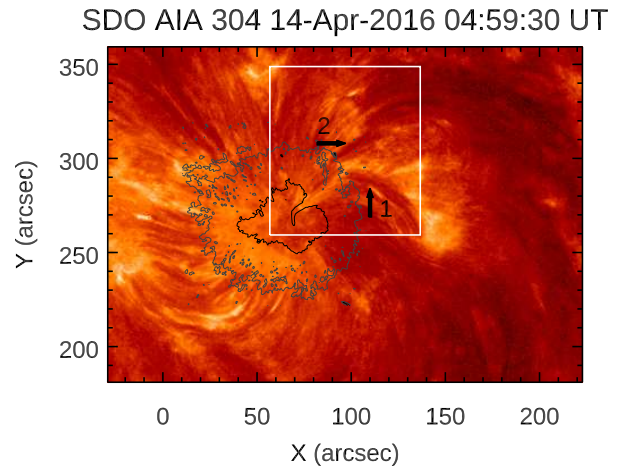}
	\caption{Image of the leading sunspot of AR~12529 as seen at 304~\AA{} by \textit{SDO}/AIA. Arrows indicate two bundles of loops (see the main text). For comparison, the contours indicate the umbra-penumbra boundary (black) and penumbra-moat boundary (grey) in the 2832~\AA{} passband and the white box frames the sub-FoV shown in Figure~\ref{fig:iris}. \label{fig:aia}}
\end{figure}

For this reason, to interpret the nature of the UF we have to take into account the observations of its counterparts in the upper atmospheric layers. High-resolution images by \textit{IRIS} are suggestive of the presence of a curved feature, with a length of about 50\arcsec, that has one edge cospatial to the UF observed in the photosphere. Both in the upper chromosphere and in the transition region, elongated bright structures together with darker patches are found to follow the same curved paths (see the dashed lines in Figure~\ref{fig:iris}). Brighter fibrils with complex motions are noted in the vicinity of the UF. 

Actually, these observations support earlier analyzed data acquired in the H$\alpha$ line with the equatorial spar at INAF -- Catania Astrophysical Observatory and at 304~\AA{} from \textit{SDO}/AIA (see Figure~3, Paper~I). To show the consistency of these results, in Figure~\ref{fig:aia} we show an image at 304~\AA{} from \textit{SDO}/AIA acquired during the first \textit{IRIS} observing sequence, simultaneous to the SJ at 2832~\AA{} displayed in Figure~\ref{fig:irisfov}. As it can be seen in these \textit{SDO}/AIA observations, a bundle of loops (``1'', indicated with an arrow) is cospatial to the elongated bright structures visible in \textit{IRIS} SJIs (Figure~\ref{fig:iris}). This spatial correspondence hints at the possibility that a flux rope, whose tail corresponds to the UF, being present in the middle of the sunspot umbra led to the formation of the UF. Moreover, \textit{IRIS} observations and \textit{SDO}/AIA images at 304~\AA{} suggest the presence of a bifurcation of field line bundles in the upper atmosphere. In fact, a filament is observed in the upper chromosphere in the \textit{SDO}/AIA image at 304~\AA{} (see the arrow pointing to ``2'' in Figure~\ref{fig:aia}). This newly considered filament is located to the north-east with respect to counterpart of the UF. Its footpoint appears being cospatial to the intense field strength area found in the inner part of the penumbral gap visible in the photosphere.


The comparison between the properties of this UF and those found in previous observations allows us to detail other interesting results. First, we highlight that the plasma motions recognized along the UF are normal Evershed flows, contrarily to those observed by \citet{Kleint:13}. Indeed, they reported a counter-Evershed flow along UFs as the most striking property of these solar features. Moreover, they found counterparts of the UFs in the upper atmospheric layers, as extended structures visible in chromospheric and coronal images going from the umbra to outside the penumbra. To explain the observed phenomena, \citet{Kleint:13} presented two schematic scenarios. In the first (\textit{sheet model}), the bright filament ending in the umbra is formed by a sheet cutting several atmospheric layers and producing a siphon inflow towards the sunspot by the pressure difference between the umbra and the network. In the second scenario (\textit{thick flux tube model}), they interpreted the emission as an effect of the increase of the optical depth in a high dense magnetic flux tube, so that the observed region is formed higher in the atmosphere. \citet{Siu-Tapia:17}, for their part, contrasted the strong counter-Evershed flow observed in a large penumbral sector to the scenarios proposed by \citet{Kleint:13}. They noticed that these flows characterized the penumbral sector in the low photosphere as well, which contradicts the hypothesis that they occur as siphon flows at chromospheric heights. However, they pointed out that the penumbral sector could be the result of penumbral formation around a pore near the main sunspot that, later, adopted the penumbral filaments when the pore disappeared. This configuration could also explain the extremely strong magnetic field strength, up to $\sim 7000$~G, found at the heads of these filaments, which would correspond to the tails of originally formed penumbral filaments where downflows may enhance the magnetic field strength. Actually, such a configuration has been recently indicated as a possible explanation for super-strong magnetic fields in sunspots \citep{Okamoto:18}. However, note that this topology is different from that we found in the UF, where the strongest magnetic field patches are characterized by a large horizontal component of the magnetic field.

In accordance with the evidence provided, the scenario proposed in Paper~I to interpret the UF as the photospheric counterpart of an overlying flux rope touching the umbra is strongly supported by the present analysis of high-resolution multi-layer observations. Cospatial to the UF, a filamentary bundle is observed in the higher atmospheric layers. Laying down on the umbra, this structure brings strong, inclined fields above it. Such magnetic configuration can trigger penumbral-like magneto-convection in the photosphere. This scenario bears some resemblance with the \textit{sheet model}. In fact, a bifurcation of field line bundles is observed in the upper atmosphere. This topological feature appears to touch the umbra and span several atmospheric layers, while connecting the different flux systems to spatially separated regions. However, siphon flow from outside of the sunspot to the umbra, leading to counter-Evershed flows, is not observed in the UF, neither turbulent motions at the boundaries of the sheet. The \textit{thick flux tube model} is also somewhat reminiscent of a flux-rope topology but, again, inverse Evershed flows, as commonly observed in the upper atmospheric levels, are not observed in the UF studied in this paper. Furthermore, the long lifetime of the UF is difficult to explain with such a model, as the massive flux tube should be dense enough to keep its structure for about 5 days. On the other hand, both of models can account for the opposite polarity seen in the structure, because of their separate magnetic topology with respect to the hosting umbra. In addition, according to these models, flows occurring along the filament ``2'' at photospheric level could explain the presence of the elongated blue-shifted patch observed in the penumbral gap region.

Indeed, the presence of stable, mostly horizontal magnetic fields has been linked to the onset of radiatively driven penumbral magneto-convective mode in different environments, if they are strong enough \citep{Jurcak:14,Jurcak:17}. Such a mechanism has been invoked to explain the formation of regular penumbrae in sunspot \citep{Shimizu:12,Romano:13,Romano:14}, also when the penumbra formation occurs in the region of ongoing flux emergence \citep{Murabito:17,Murabito:18}, as confirmed by numerical simulations \citep[e.g.][]{Rempel:12,MacTaggart:16}. Moreover, it seems to account for the formation of other penumbral-like structures, such as orphan penumbrae, when the emerging flux is overarched by an overlying canopy \citep{Lim:13,Guglielmino:14,Zuccarello:14}, but also when they appear as photospheric manifestations of flux ropes trapped highly in the photosphere, corresponding to chromospheric filaments \citep[e.g.,][]{Kuckein:12a,Kuckein:12b,Buehler:16}. Hence, the nature of the UF can be explained in terms of an external strong horizontal field leading to the activation of the penumbral magneto-convective mode within the sunspot umbra. 

A possible origin for this structure might be related to a magnetic field rearrangement occurring in the AR. Changes in the magnetic field topology have been recently invoked to explain the formation of penumbral filaments also in mature spots \citep{Louis:13,Verma:18,Romano:19}. The overlying flux rope could stabilize a configuration where strong horizontal fields are present for a quite long time (some days), explaining the lifetime of the observed UF.

A further confirmation of this scenario comes from the fact that many pixels in the region of the UF need a double-component inversion to be properly modeled. We tried to fit these peculiar profiles with several complex inversion schemes, which involve different kinds of stratification along the optical depth of the magnetic field strength, LOS velocity, inclination and azimuth angles. Among those, we adopted the double-component inversion scheme with the smaller $\chi^{2}$ merit function. In the Appendix, we show that in the UF region the flux-tube component has a stronger magnetic field intensity. The distribution of the relative filling factor between the two components is similar to that in the penumbra. This suggests that a large penumbral-like filament, being the origin of the flux-tube component, is hanging above the sunspot umbra, which is responsible for the umbral component. 
The choice of such a complex inversion scheme allows us the determination of physical quantities with a high accuracy around optical depth $\tau_{500}=0.1-0.01$ but produces inaccurate results at deeper layers. This could be alleviated using additional diagnostics such as the infrared pair of \ion{Fe}{1} lines at 1565~nm \citep[see for instance][]{Borrero:16}. Even so, our lines provide an information essential to understand the nature of the UF: that is, a flux system characterized by a strong horizontal field sharing the same area with an atmosphere representing the ``classical'' sunspot umbra. This is evident in the distribution of the horizontal field strength in the UF (see the Appendix).
	
In this regard, it must be noted that such a configuration is reminiscent of the two components -- spines/intra-spines -- observed in regular penumbrae: the horizontal component, which is referred to as intra-spines, or flux-tube component, that carries the Evershed flow, and other more vertical component, spines or background component, with stronger magnetic field intensity. Such a fluted or ``uncombed'' structure of the penumbral magnetic fields on small scales along the azimuthal direction was recognized in observations by \citet{Lites:93,Solanki:93,Title:93} and later confirmed by a number of investigations with increasing spatial and spectral resolution (\citealp{Pillet:00,Schlichenmaier:02,Bellot:04}; more recently, \citealp{Tiwari:15} and \citealp{Murabito:19}, who found such a fine structure also at the chromospheric level).

However, in our observations we find that in the UF the umbral component, which would correspond to the spines component, characterized by the sunspot polarity and usually vertical, has often a weaker field strength than the horizontal, flux-tube (or penumbral) component, which would represent the intra-spines. This result prevents us to identify the vertical and horizontal components with the spines and intra-spines, respectively. Multi-line diagnostics at different heights is necessary to clarify this issue. Furthermore, the UF core exhibits Stokes profiles alike those seen in the regular penumbra (compare the blue and red profiles in Figure~\ref{fig:profiles}), except for the sign of the polarity, even considering a single atmospheric component. Therefore, this argument supports an interpretation of the UF as a structure with strong horizontal fields (i.e., a flux rope) overlying the sunspot umbra. The possibility that this structure is indeed embedded in the umbra, cutting it, could be considered an alternative explanation. Nevertheless, we cannot rule out that a single-component inversion with a more complex atmosphere stratification could be able to model the observed Stokes profiles also in those pixels of the periphery of the UF, although this solution might appear inadequate taking into account the scenario deduced from the overall analysis. 

Additional observational facts, supporting the hypothesis that the UF may be due to the presence of a flux rope sinking in the middle of the sunspot, are the spatial distribution of the inclination and azimuth angles along the UF. Their symmetry along the axis of the UF can be interpreted as a consequence of a convective pattern similar to that found in the model proposed by \citet{Borrero:10}. The observational properties that are in agreement with our observations are naturally explained in terms of penumbral structures, regardless their origin. In fact, in a penumbral-like structure determined through magneto-convection by an external flux system, with different (i.e., opposite) orientation from the sunspot field, the overturning convection would give rise to the opposite polarity along the axis of the UF and to the same polarity along the lanes, like in these observations. 

In this context, one may investigate on the origin of brightenings observed along the UF, provided that this intruding structure could interact with the main flux systems of the sunspot. The observations reported by \citet{Barthi:17,Barthi:18} show jets and brightenings in the chromosphere and transition region above filamentary structures intruding sunspot umbrae. These phenomena have been interpreted as a result of magnetic reconnection taking place between the umbral field and the opposite-polarity patches observed along the penumbral-like intrusions, due to strong convective motions that enhance or even reverse the magnetic field. However, the region studied by \citet{Barthi:18} was observed in a $\delta$ spot, where the uncombed configuration of the penumbrae and the shear present in the interface region between the two opposite-polarity umbral cores may give rise to a complex magnetic and velocity field configuration \citep[e.g.,][]{Cristaldi:14,Shimizu:14}. We cannot exclude that such interactions may occur in the system we analyzed, although one has to keep in mind that AR~12529 was not flare-productive at all, in contrast to ARs studied by \citet{Kleint:13} and \citet{Barthi:18}.

Nevertheless, in \textit{IRIS} images a compact bright knot is found, cospatial to the penumbral gap in the photosphere and close to the direction along which the elongated brightenings are observed. Actually, this seems to correspond with the footpoint of the filament ``2''. Magnetic reconnection occurring between these two different flux systems 
in the vicinity of the bright footpoint could explain why we observe fibrils with enhanced emission in this region along the UF.

All the same, it should be emphasized that this is only a possible interpretation, mainly based on the spectropolarimetric measurements. The presence of a multi-component atmosphere within the same pixels belonging to the UF, or rather to a complex atmospheric stratification, indicates that it is very difficult to determine the magnetic configuration of this feature. In addition, the way we dealt with the $180^{\circ}$-azimuth ambiguity through the comparison with a potential extrapolated field may be inaccurate in the region of the UF. 

The hypothesis about the nature and origin of UFs could be confirmed or ruled out by a careful analysis of future observations of similar structures. New measurements, comprising both higher spatial resolution and multi-layer polarimetric coverage, are needed to properly disentangle the signals coming from different flux systems. In this perspective, we expect that this conundrum has to wait until the next generation of large-aperture telescopes, like the DKIST \citep[Daniel K.~Inouye Solar Telescope,][]{Keil:10} and EST \citep[European Solar Telescope,][]{Collados:10}, to be solved.

\acknowledgments

The authors would like to thank the anonymous referee for his/her insightful comments and suggestions. 
This work was supported by the Istituto Nazionale di Astrofisica (PRIN-INAF 2014), by the Italian MIUR-PRIN grant 2012P2HRCR on The active Sun and its effects on space and Earth climate, by Space Weather Italian COmmunity (SWICO) Research Program, and by the Universit\`a degli Studi di Catania (Piano per la Ricerca Universit\`{a} di Catania 2016-2018 -- Linea di intervento~1 ``Chance''; Linea di intervento~2 ``Ricerca di Ateneo - Piano per la Ricerca 2016/2018''). The research leading to these results has received funding from the European Union's Horizon 2020 research and innovation programme under grant agreement no.~739500 (PRE-EST project) and no.~824135 (SOLARNET project). This work has been partially funded by the Spanish Ministry of Economy and Competitiveness through the Project no.~ESP-2016-77548-C5. \textit{Hinode} is a Japanese mission developed and launched by ISAS/JAXA, with NAOJ as domestic partner and NASA and STFC (UK) as international partners. It is operated by these agencies in co-operation with ESA and Norwegian Space Centre. \textit{IRIS} is a NASA small explorer mission developed and operated by LMSAL with mission operations executed at NASA Ames Research center and major contributions to downlink communications funded by ESA and the Norwegian Space Centre. \textit{SDO}/HMI data used in this paper are courtesy of NASA/\textit{SDO} and the HMI science team. Use of NASA's Astrophysical Data System is gratefully acknowledged.

\facility{\textit{Hinode} (SOT), \textit{IRIS}, \textit{SDO} (HMI/AIA)}

\appendix

\begin{figure}[t]
	\centering
	\includegraphics[trim=0 0 30 0, clip, scale=0.75]{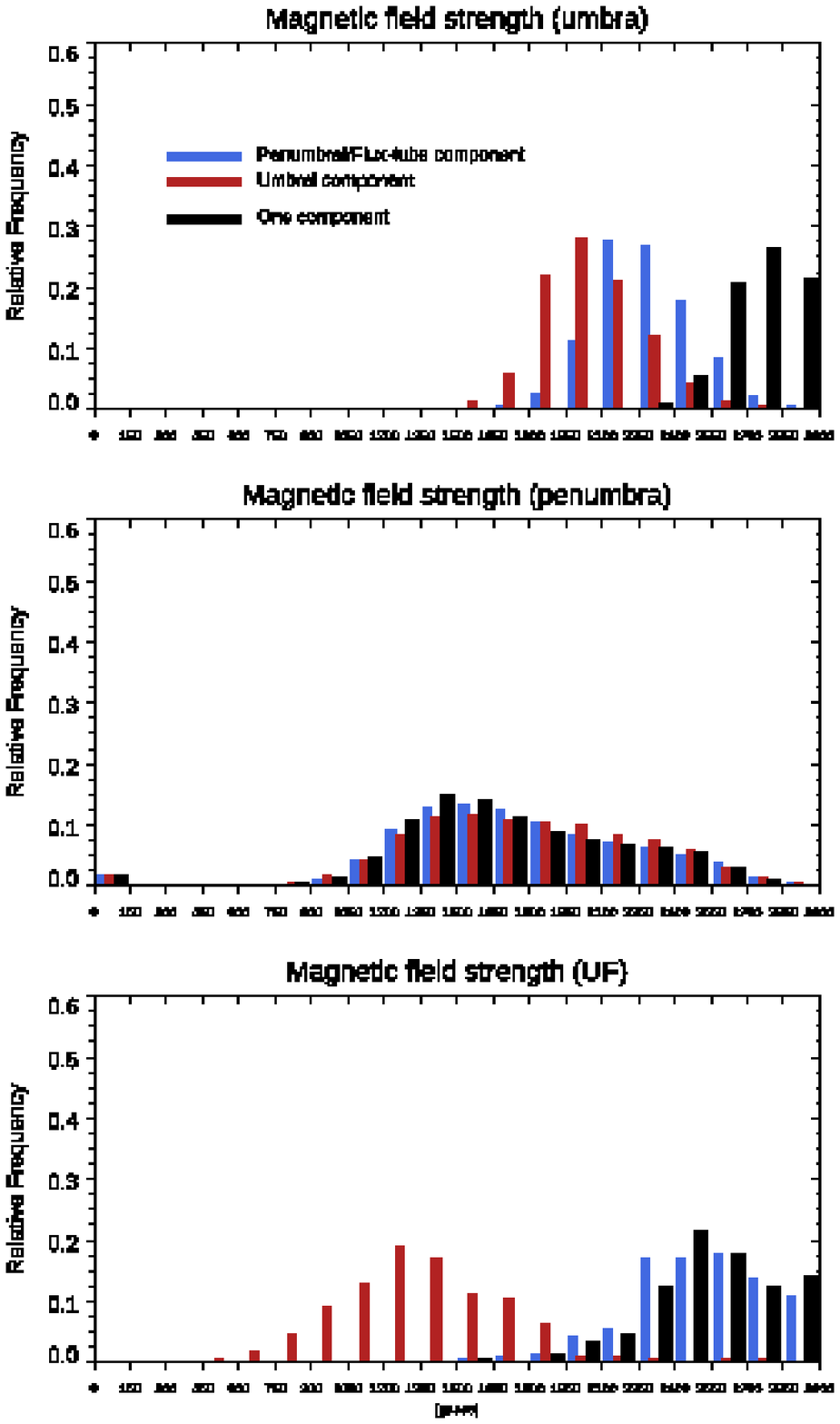}%
	\includegraphics[trim=40 0 30 0, clip, scale=0.75]{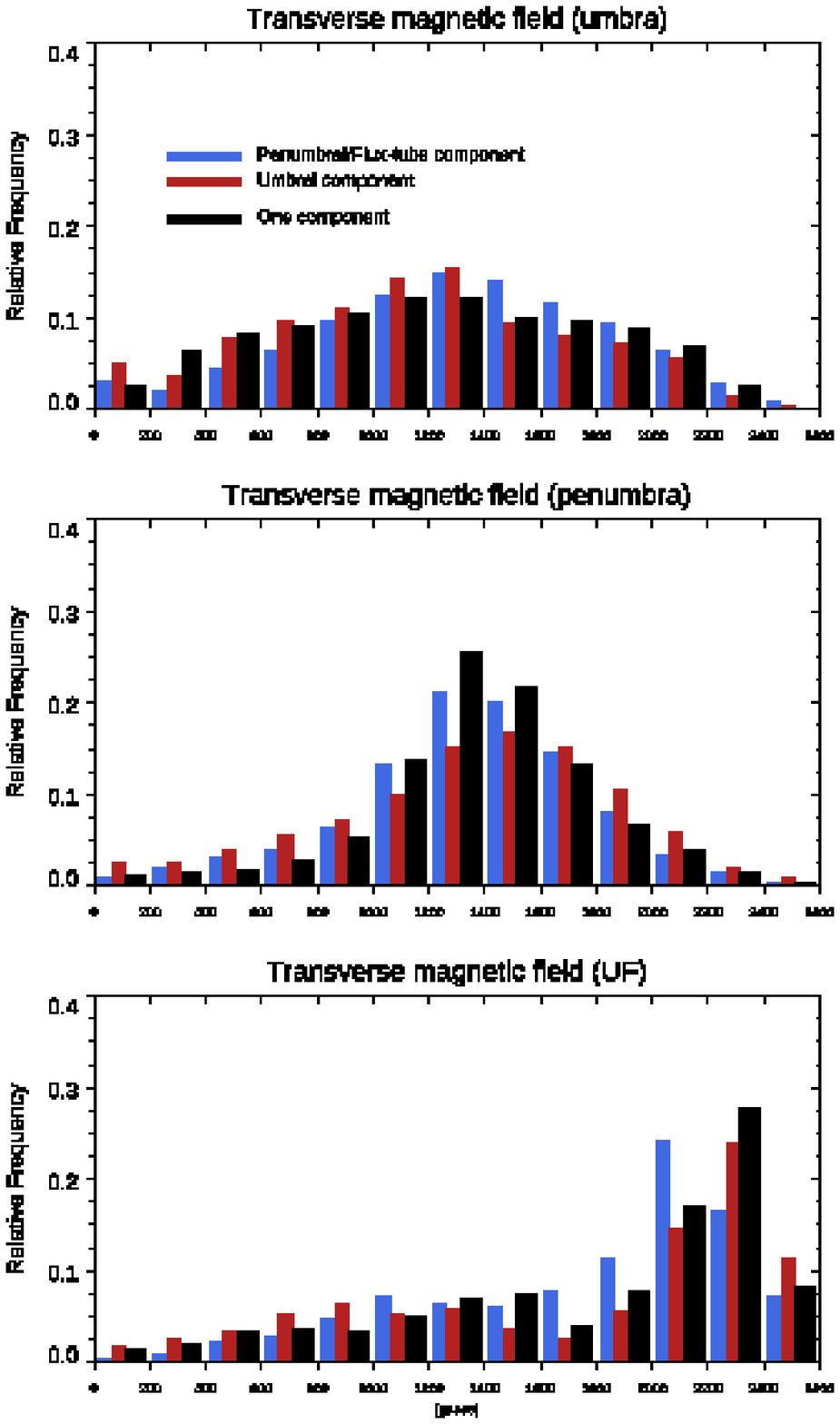}
	\caption{Histograms relevant to the magnetic field strength (left) and to the horizontal component of the magnetic field (right) for the three regions of the FoV. \label{fig:magfil}}
\end{figure}

\begin{figure}[t]
	\centering
	\includegraphics[trim=0 0 30 0, clip, scale=0.75]{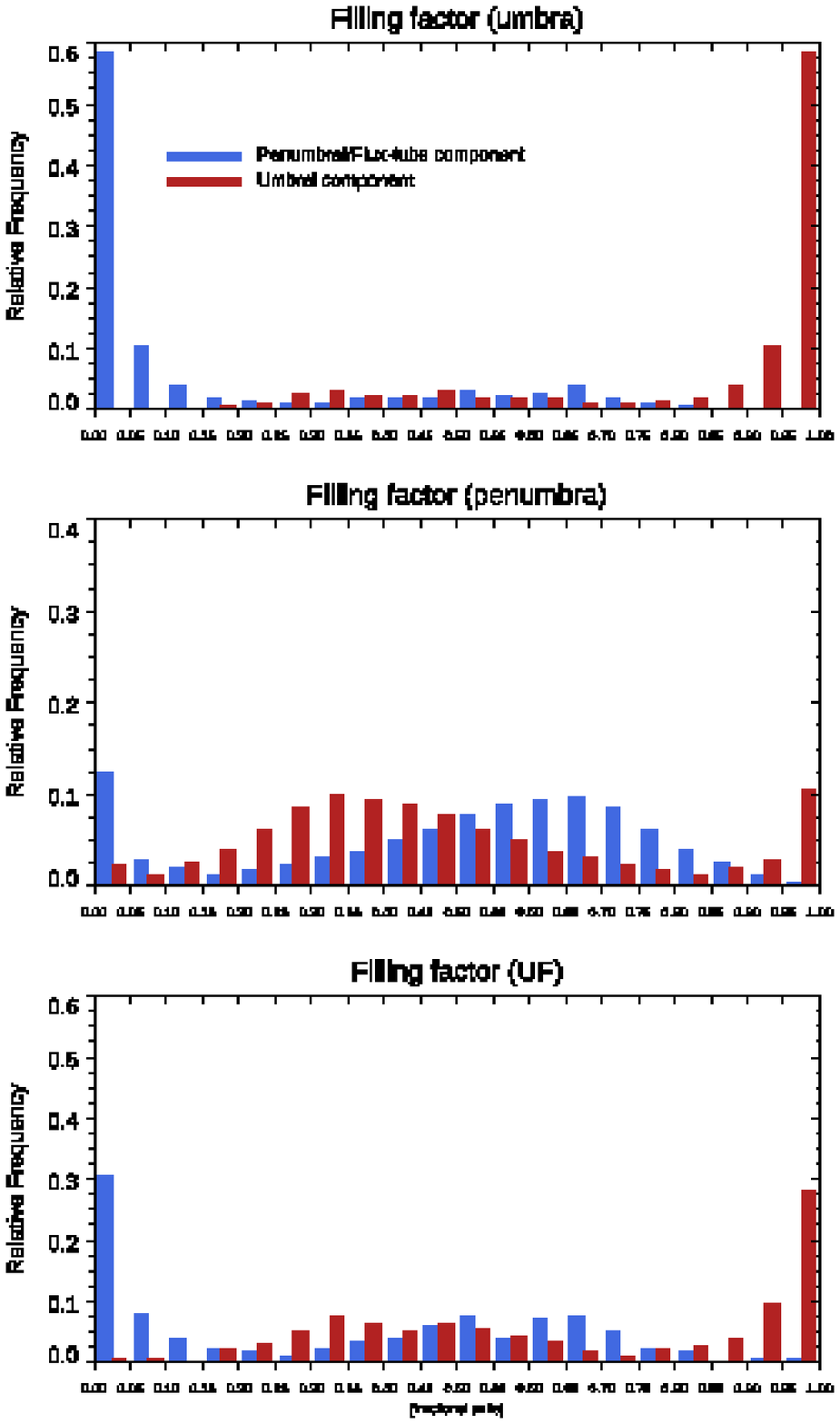}%
	\includegraphics[trim=40 0 30 0, clip, scale=0.75]{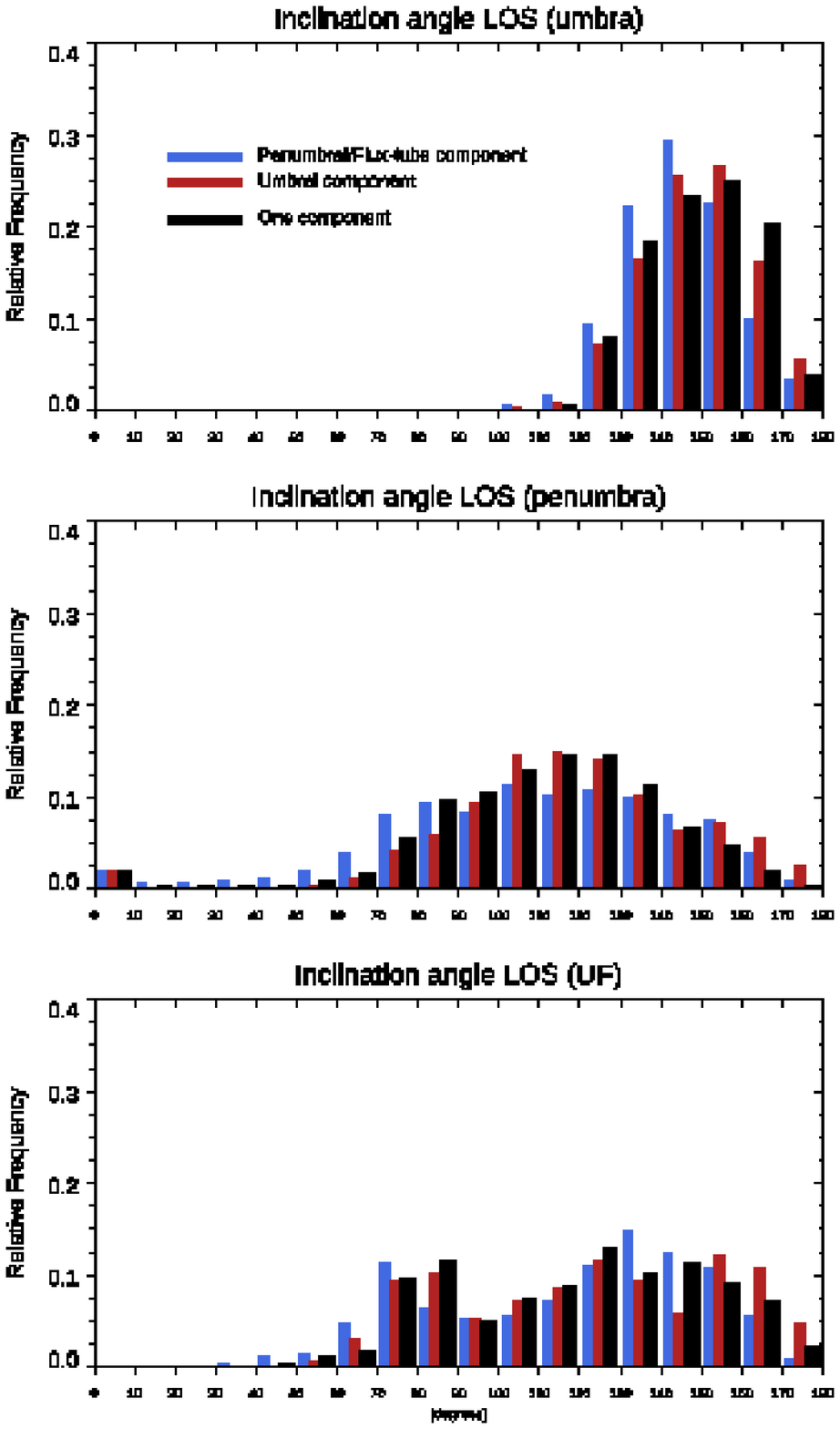}
	\caption{Same as in Figure~\ref{fig:inctra}, for the relative filling factor (left) and for the inclination angle (right). \label{fig:inctra}}
\end{figure}

In order to make a comparison between the single- and double-component inversions, we have identified three regions in the FoV: umbra, penumbra, and UF. Umbra and penumbra have been selected by using a threshold in the continuum intensity, the UF has been drawn by hand. We plot in Figures~\ref{fig:magfil} and~\ref{fig:inctra} the histograms relevant to the distribution of the two components, identified as ``penumbral'' (or ``flux-tube'', blue color) and ``umbral'' (red color) components, in the three regions of the FoV, compared to that retrieved by the single-component inversion (black color). We have evaluated all the physical quantities at $\log \tau_{500} = -1.5$.

Figure~\ref{fig:magfil} (left panel) indicates that the penumbral component in the UF region has a higher field strength than the umbral one. Figure~\ref{fig:magfil} (right panel) clearly shows that the UF has a strong horizontal component of the vector magnetic field, as both components peak at high values of about $2000-2500$~G.

The distribution of the relative filling factor (Figure~\ref{fig:inctra}, left panel) \textbf{in the UF} has some resemblance to the case of the penumbra, with a slight prevalence of the penumbral component at intermediate values. 

As regards the inclination in the UF (Figure~\ref{fig:inctra}, right panel), the penumbral component exhibits a nearly bimodal distribution of the inclination angles, with one peak at $\approx 75^{\circ}$, representing a small part of the pixels which are mostly horizontal, and a larger fraction of the pixels that tend to have inclination angles distributed around $135^{\circ}$. On the other hand, the umbral component has an extended tail towards inclination angles of about $150^{\circ}$. Note that these results are relative to the LOS reference frame.

\end{document}